\documentclass{aa}  

\usepackage{graphicx}
\usepackage[caption=false,font=footnotesize]{subfig}
\usepackage{color}
\usepackage{textcomp}
\usepackage[varg]{txfonts}
\usepackage{natbib}
\usepackage{lscape}
\usepackage{wasysym}
\usepackage{gensymb}
\usepackage{txfonts}
\usepackage{natbib}
\bibpunct{(}{)}{;}{a}{}{,} 
\newcommand\refbold[1]{#1}

\newcommand\refboldd[1]{#1}

\begin{document}

   \title{Multi-spacecraft observations and transport simulations of solar energetic particles for the May 17th 2012 event}
   \titlerunning{Multi-spacecraft observations and transport simulations of solar energetic particles}

   \author{M. Battarbee
          \inst{1}\thanks{Currently at the Department of Physics, University of Helsinki, Finland}
          \and
          J. Guo
          \inst{2}
          \and
          S. Dalla
          \inst{1}
          \and
          R. Wimmer-Schweingruber
          \inst{2}
          \and
          B. Swalwell
          \inst{1}
          \and
          D. J. Lawrence
          \inst{3}  
          }

   \institute{Jeremiah Horrocks Institute,
             University of Central Lancashire, PR1 2HE, Preston, UK \\
         	 \email{markus.battarbee@gmail.com}
		 \and	
         	 Institut fuer Experimentelle und Angewandte Physik,
			 University of Kiel, Germany \\
	         \email{guo@physik.uni-kiel.de}
         \and 
             Johns Hopkins University Applied Physics Laboratory, MD, USA
             }

   \date{Received 26th June 2017 / Accepted 25th January 2018}

 
  \abstract
   {
The injection, propagation and arrival of solar energetic particles (SEPs) during eruptive solar events is an important and current research topic of heliospheric physics. During the largest solar events, particles may have energies up to a few GeVs and sometimes even trigger ground-level enhancements (GLEs) at Earth. These large SEP events are best investigated through multi-spacecraft observations.}
   {We study the first GLE-event of solar cycle 24, from 17th May 2012, using data from multiple spacecraft (SOHO, GOES, MSL, STEREO-A, STEREO-B and MESSENGER). These spacecraft are located throughout the inner heliosphere, at heliocentric distances between 0.34 and 1.5 astronomical units (au), covering nearly the whole range of heliospheric longitudes.}
   {We present and investigate sub-GeV proton time profiles for the event at several energy channels, obtained via different instruments aboard the above spacecraft. We investigate issues due to magnetic connectivity, and present results of three-dimensional SEP propagation simulations. We gather virtual time profiles and perform qualitative and quantitative comparisons with observations, assessing longitudinal injection and transport effects as well as peak intensities.}
   {We distinguish different time profile shapes for well-connected and weakly connected observers, and find our onset time analysis to agree with this distinction. At select observers, we identify an additional low-energy component of Energetic Storm Particles (ESPs). Using well-connected observers for normalisation, our simulations are able to accurately recreate both time profile shapes and peak intensities at multiple observer locations.}
   {This synergetic approach combining numerical modelling with multi-spacecraft observations is crucial for understanding the propagation of SEPs within the interplanetary magnetic field. Our novel analysis provides valuable proof of the ability to simulate SEP propagation throughout the inner heliosphere, at a wide range of longitudes. Accurate simulations of SEP transport allow for better constraints of injection regions at the Sun, and thus, better understanding of acceleration processes.}


   \maketitle
%

\section{Introduction}
The Sun releases vast amounts of energy through its activity, which mostly follows a periodic 11-year cycle. These eruptions can accelerate protons, electrons and heavier ions to relativistic energies and release them into interplanetary space. These solar energetic particles (SEPs) are guided by the interplanetary magnetic field (IMF), and in some cases result in intensive particle fluxes near the Earth. SEP events take place more frequently during solar maximum, and can affect atmospheric and space-related activities in many ways \refbold{(see, e.g., \citealt{Turner2000} and references therein)}, and as such, their investigation has been recognized as extremely important.

During extreme solar events, protons can be accelerated into the GeV range, and, when directed at the Earth, may lead to neutron monitors (NMs) detecting events at the Earth's surface. These ground-level enhancements (GLEs) are the most extreme of solar events, and thus are of special interest to the heliophysics community \refbold{(see, e.g., \citealt{Asvestari2017} and \citealt{Nitta2012})}. Our understanding of energetic solar events and specifically GLEs increased dramatically during solar cycle 23 \citep{Gopalswamy2012} due to advances in instrumentation and an abundance of events to observe. Solar cycle 24, being much quieter, has so far provided only 
\refboldd{two unambiguous GLEs, GLE71 on May 17th 2012 and GLE72 on September 10th 2017. }
\refbold{Being able to observe this event from multiple vantage points within the inner heliosphere provides us with an exciting opportunity to increase our understanding the dynamics of strong solar events. In such an analysis, three-dimensional modelling of particle propagation is a crucial tool.}

We present sub-GeV proton observations of GLE 71, \refbold{focusing on comparative analysis between observations from} multiple vantage points throughout the inner heliosphere to better understand the spatial extent of SEP intensities in \refbold{strong SEP events. GeV-energy particles are thus excluded from our analysis due to observations at such energies being available only in the vicinity of the Earth.} We present new observations from the Mars Science Laboratory (MSL) Radiation Assessment Detector (RAD) and the MESSENGER Neutron Spectrometer (NS), together with energetic particle data from STEREO and near-Earth missions. We use a fully three-dimensional test particle model to simulate the transport of SEPs, originating from an acceleration region in the solar corona, generating virtual time profiles at various observer locations. The model includes, for the first time, the effects of a wavy Heliospheric Current Sheet (HCS) \refbold{between} two opposite polarities of the IMF. We compare intensity time profiles and peak intensities of data from both observations and simulations, at the different observer locations.

In section \ref{subsec:gle71}, we introduce the event along with previously published analysis. In section \ref{sec:observations}, we introduce the instruments used in our multi-spacecraft observations. We then present intensity time profiles and solar release times, and discuss magnetic connectivity and energetic storm particles (ESPs). In section \ref{sec:simulations}, we describe our particle transport simulation method. We then proceed to present simulated intensity time profiles, and compare them and deduced peak intensities with observations. Finally, in section \ref{sec:conclusions} we present the conclusions of our work. In Appendix \ref{appendix:mes}, we discuss calibration of our MESSENGER NS observations.

\section{The May 17th 2012 solar eruption}\label{subsec:gle71}

On May 17, 2012 at 01:25 UT, the NOAA active region 11476, located at \mbox{N11 W76} in Earth view, produced a class M5.1 flare starting, peaking, and ending at 01:25, 01:47, and 02:14 UT, respectively \citep[e.g.,][]{Gopalswamy2013, Shen2013}.
The type II radio burst indicating the shock formation was reported by \citet{Gopalswamy2013} to start as early as 01:32 UT using the dynamic spectra from Hiraiso, Culgoora and Learmonth observatories.
Based on this, they also determined the coronal mass ejection (CME) driven shock formation height as 1.38 solar radii ($R_\odot$, from the centre of the Sun). The CME reached a peak speed of \mbox{$\sim$ 1997 km s$^{-1}$} at 02:00 UT. 
They reasoned that although the May 17th flare is rather small for a GLE event, the associated CME was directed toward near-ecliptic latitudes, facilitating good connectivity between the most efficient particle acceleration regions of the shock front and the Earth. Despite the flare exhibiting relatively weak x-ray flux, \cite{Firoz2014,Firoz2015} suggested that both the flare and the CME had a role in particle acceleration. \cite{Ding2016} agreed with this, based on velocity dispersion analysis (VDA) of proton arrival.

\citet{Gopalswamy2013} further estimated, using NM data, that the solar particle release time was about 01:40, slightly later than the shock formation time of 01:32. \citet{Papaioannou2014} reported the type III radio bursts which signified the release of relativistic electrons into open magnetic field lines starting at around 01:33 UT and ending at 01:44 UT. 
Using a simple time-shifting analysis, they derived the release of \mbox{1 GeV} protons from the Sun at about 01:37 UT, slightly earlier but broadly agreeing with the onset time obtained by \citet{Gopalswamy2013}.   

This event was later directly detected at Earth by several NMs\footnote{\url{http://www.nmdb.eu}} with slightly different onset times (between 01:50 and 02:00), with the strongest signal detected at the South Pole \citep{Papaioannou2014} where the rigidity cutoff is the lowest.
Within the magnetosphere, proton energy spectra were measured by the 
PAMELA instrument \citep{picozza2007pamela} as reported by \citet{adriani2015pamela}, indicating that protons with energies of up to one GeV and helium of up to 100 MeV/nucleon were accelerated and transported to the vicinity of Earth. The GeV proton detection has also been corroborated later by \citet{kuhl2015} using an inversion technique exploring the response functions of the Electron Proton Helium Instrument \citep[EPHIN,][]{Muller-Mellin1995} aboard the SOHO spacecraft. The event was also detected aboard the international space station \citep{Berrilli2014}.
Analysis of NM and PAMELA observations, using comparisons of peak and integral intensities, can be found in \cite{Asvestari2017}. \refbold{\cite{Mishev2014} performed reverse modelling based on NM measurements of this event, finding evidence of anisotropic twin-stream SEP pitch-angle distributions.}

Utilizing lower particle energies for release time analysis, \cite{Li2013} compared Wind/3DP and \mbox{GOES 13} particle fluxes with NM and solar disk observations, concluding that electrons at this event appear to be flare-accelerated, with proton acceleration happening mainly at the CME-driven shock. The ERNE/HED detector \citep{Torsti1995} aboard SOHO detected a strong event, but suffered from data gaps during the event, which poses additional challenges to analysis.

During this event, the STEREO Ahead (STA) and STEREO Behind (STB) spacecraft were leading and trailing Earth by 114.8 and 117.6 degrees, respectively, both at a heliocentric distance of approximately \mbox{1 au}.
\citet{lario2013} studied the \mbox{15-40 MeV} and \mbox{25-53 MeV} proton channels of this event using GOES and the high energy telescope (HET) on STB. 
For the \mbox{15-40 MeV} channel, they obtained an enhancement rate (peak intensity/pre-event intensity) of \mbox{2.64 $\times 10^3$} at GOES and only 35.0 at STB.  
For the \mbox{25-53 MeV} channel, they obtained an enhancement rate of \mbox{1.94 $\times 10^4$} at GOES and only 13.4 at STB.  
Unfortunately they did not determine the peak intensity of this event as measured by STA. 
This event has previously been included in a STEREO event catalogue \citep{Richardson2014}, and multi-spacecraft observations of electrons have been analysed in \cite{Dresing2014}. \cite{Heber2013} included STA and STB proton time profiles for a single energy range in a figure, displaying the longitudinal extent of the event.

The event was also observed by the MESSENGER (MES) spacecraft orbiting around Mercury which, at the time of the event, was at a heliocentric distance of \mbox{0.34 au} \citep{Lawrence2016}. The longitudinal connectivity of MES was similar to that of STA, as shown in Figure \ref{fig:heliospheric_ps}. In this paper, we investigate the time-series of proton measurements from MES using its neutron spectrometer \citep[NS,][]{Lawrence2016}. 

Beyond \mbox{1 au}, this event was also observed by the Radiation Assessment Detector \citep[RAD,][]{hassler2012} on board the Mars Science Laboratory (MSL) on its way to Mars \citep{zeitlin2013}. 
We derive the proton intensities measured by RAD at different energy ranges and compare them with Earth-based observations and simulated particle intensities at the same location. 
We note that the RAD detector did not measure original proton intensities in space, but rather a mix of primary and secondary particles due to primaries experiencing nuclear and electromagnetic interactions as they traverse through the inhomogeneous flight-time shielding of the spacecraft. To retrieve the original particle flux outside the spacecraft is rather challenging and is beyond the scope of the current paper. 

\section{Multi-spacecraft observations}\label{sec:observations}

\begin{figure}[htb!]
{\includegraphics[trim=80 200 20 200,clip, scale=0.45]{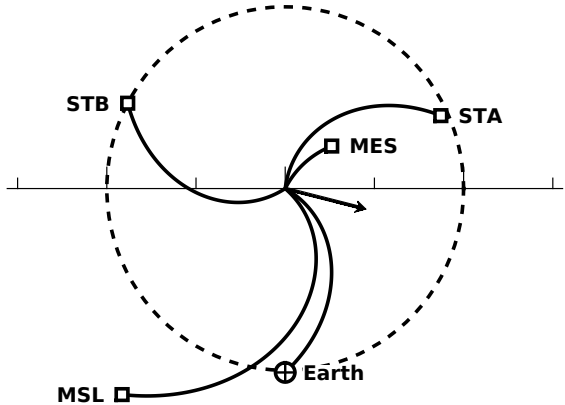}}
\caption{The heliospheric locations of MES, Earth, MSL, STA and STB. The Parker spiral configuration is calculated using a constant solar wind speed of $u_\mathrm{sw}=400\,\mathrm{km}\,\mathrm{s}^{-1}$. The \mbox{1 au} distance is shown with a dashed circle. The arrow is placed along the radial direction at the flare location.}\label{fig:heliospheric_ps}
\end{figure}

The heliospheric locations of five different spacecraft whose measurements are employed in the current study are shown in Figure \ref{fig:heliospheric_ps} and also listed in Table \ref{tab:heliospheric_ps}. For this study, we estimated the average solar wind speed from measurements made by the CELIAS/MTOF Proton Monitor on the SOHO Spacecraft during Carrington rotation 2123. The average radial solar wind speed value was \mbox{410 km s$^{-1}$}, which was rounded down to \mbox{400 km s$^{-1}$} for the purposes of this research. Table \ref{tab:heliospheric_ps} also includes calculated Parker spiral lengths using this solar wind speed.

\begin{table*}{}
\caption{Heliospheric location, Parker spiral length and onset time of the event seen at different spacecraft. The flare source region at the Sun is NOAA active region 11476 with coordinate of N11 W76 and the flare onset time is 01:25 on 17th May 2012.} \label{tab:heliospheric_ps}
\begin{tabular}{crrrr|r|r|r|r }     
 & HGI & HGI & distance   & Parker spiral & shortest & SEP & estimated SRT & Observed SEP  \\
  &  latitude       & longitude       & to the Sun & length        & travel time & onset time & (1 GeV p) & event type \\
\hline
STA & 7.3 \textdegree  & 275.4 \textdegree & 0.96 au & {1.11 au} & 631.4 s & 10.18 & 10:07 & slowly rising \\
MES & 2.1 \textdegree  & 290.9 \textdegree & 0.35 au & {0.36 au} & 204.8 s & 03:14 & 03:11 & slowly rising \\
Earth & -2.4 \textdegree & 160.7 \textdegree & 1.01 au & {1.18 au} & 671.2 s & 01:56 & 01:45 & rapidly rising\\
MSL & -7.3 \textdegree & 121.8 \textdegree & 1.46 au & {1.92 au} & 1092 s & 02:04 & 01:46  & rapidly rising\\
STB & -4.7 \textdegree & 42.7 \textdegree & 1.00 au & {1.16 au} & 659.8 s & 11:06  & 10:55 & slowly rising \\
\hline
\end{tabular}
\end{table*}

\begin{figure}[htb!]\includegraphics[trim=20 50 20 50,clip, scale=0.55]{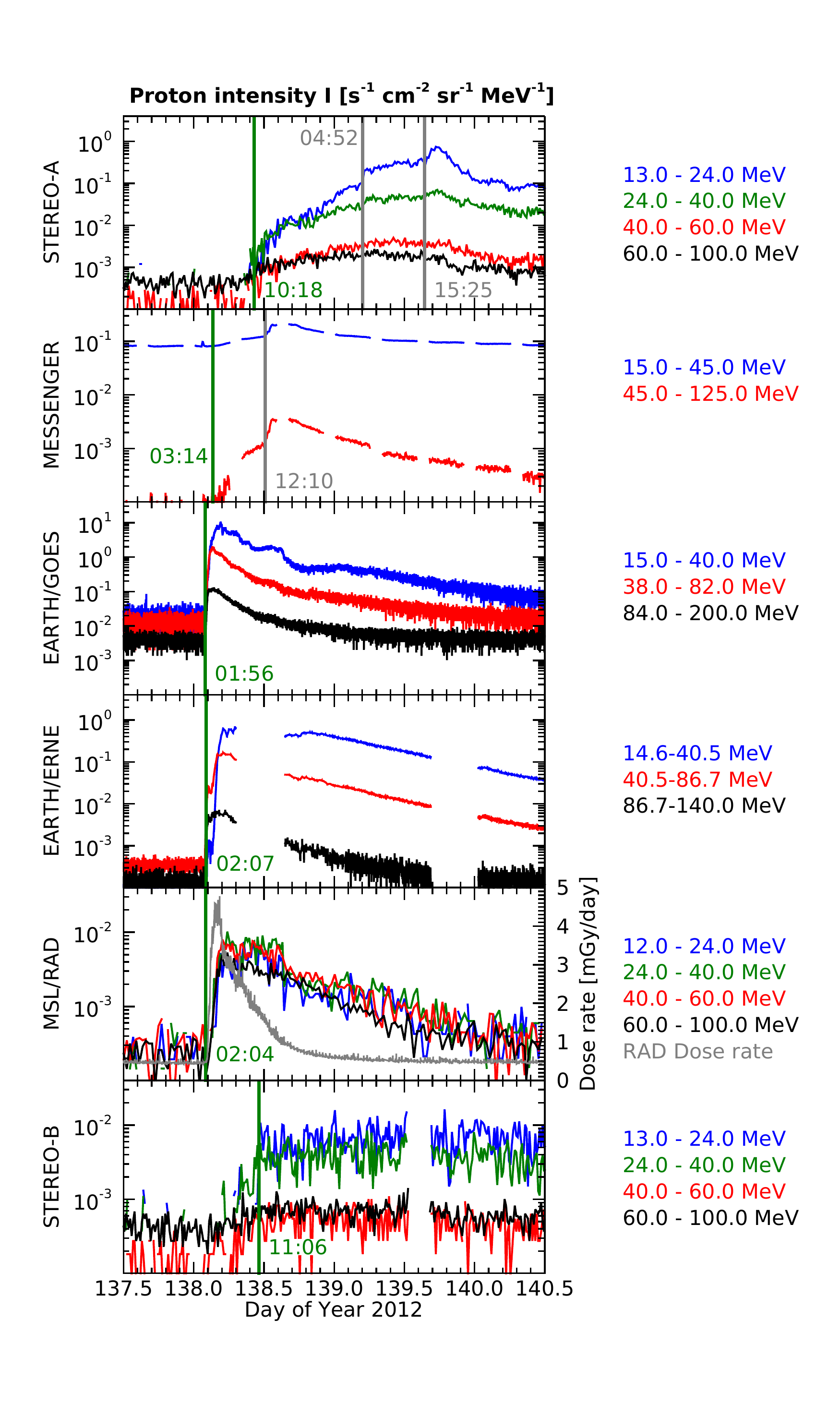}
\caption{The proton intensity time profiles, in units $\mathrm{s}^{-1}\,\mathrm{cm}^{-2}\,\mathrm{sr}^{-1}\,\mathrm{MeV}^{-1}$, for different proton energy ranges at various spacecraft. The green vertical lines mark the onset times of the first arriving particles while the grey vertical lines mark the possible onsets of ESP events. SOHO/ERNE has two large data gaps but is located close to GOES, allowing cross-comparison of the data. The 17th of May is DOY 138.}\label{fig:flux}
\end{figure}

In order to effectively analyse the heliospheric and temporal extent of the May 17th 2012 \refbold{solar eruption}, we assess proton time profiles from multiple instruments throughout the inner heliosphere. The energy-dependent time profiles of SEPs measured at five different heliospheric locations are shown in Figure \ref{fig:flux}. 

For STA and STB, we analyse 1-minute resolution data from HET of the In situ Measurements of Particles and CME Transients (IMPACT) investigation aboard both STEREOs. The protons are measured between 13 and 100 MeV in 11 different energy channels. For our purpose of comparing the STEREO measurement to those at other locations, we combine the energy channels into four different bins: \mbox{13--24 MeV}, \mbox{24--40 MeV}, \mbox{40--60 MeV}, and \mbox{60--100 MeV}.  

For MES data at Mercury, we use the neutron spectrometer which contains one borated plastic (BP) scintillator sandwiched between two Li glass (LG)  scintillators.
To account for the shielding of particles by the magnetosphere of Mercury and by the geometric shadowing of the planet itself, we selected only observations where the orbit altitude of MES is \refbold{greater} than \mbox{5000 km}.
The energy thresholds for triggering each type of charged particle were simulated and derived using particle transport codes \citep{lawrence2014detection} and are as follows: single coincidence, $\ge$15 MeV protons (or $\ge$1 MeV electrons); double coincidences, $\ge$45 MeV protons (or $\ge$10 MeV electrons); and triple coincidences, $\ge$125 MeV protons (or $\ge$30 MeV electrons). 
Since $\ge$10 MeV electrons are fairly rare in SEPs, we assume these channels measure mainly protons during the event. For the single-coincidence channel, contamination by many different sources \refbold{such as electrons, gamma-rays and various charged particles is possible, and thus, care must be taken when drawing conclusions from the flux}. We converted single, double, and triple coincidence counts into fluxes according to methods explained in detail in Appendix \ref{appendix:mes}.

We solve the intensity profile for 15--45 MeV and 45--125 MeV protons in the following way: We subtract the $\ge$45 MeV flux from the $\ge$15 MeV flux, and the $\ge$125 MeV flux from the $\ge$45 MeV flux. These two fluxes, now bounded from both above and below in energy, are then divided with the energy bin widths, i.e., 30 and 80 MeV, resulting in intensities in units \mbox{protons s$^{-1}$ sr$^{-1}$ cm$^{-2}$ MeV$^{-1}$}. 
The $\ge$125 MeV flux is not shown in Figure \ref{fig:flux}, as it shows little enhancement for this time period.  We emphasize that the 15--45 MeV flux calibration is uncertain. The time profiles in Figure \ref{fig:flux} indeed show a very high intensity in the \mbox{15--45 MeV} channel, likely due to non-proton background contamination.


Close to \refbold{the} Earth, we employed two separate detectors. \mbox{GOES 13}, situated within the Earth's magnetosphere, provided us with \mbox{15--40 MeV}, \mbox{38--82 MeV}, and \mbox{84--200 MeV} proton channels, with 32 second resolution.
The SOHO/ERNE HED detector at L1 was used to construct 
energy channels with 1 minute time resolution, \refbold{matching} 
the GOES channels with energy ranges of \mbox{14.6--40.5 MeV}, \mbox{40.5--86.7 MeV}, and \mbox{86.7--140 MeV}. 
GOES provided uninterrupted observations 
\refbold{of the event, but the background levels were enhanced due to increased ambient particle densities in the magnetosphere.} 
ERNE/HED\refbold{, located outside the magnetosphere at the Lagrangian point L1,} provided uncontaminated \refbold{pre-event intensities}, but with data gaps during the event. \refbold{Additionally, the peak intensities observed by ERNE/HED are suspected to be incorrect due to non-linear saturation artefacts and particles propagating through the detector in the reverse direction.}


At MSL, during the cruise phase, the RAD instrument provided radiation dose measurements with a high time resolution of 64 seconds, and particle spectra with a time resolution of $\sim$32 minutes. The radiation dose measurements were used to determine the event onset time. The particle spectra are provided by a particle telescope consisting of silicon detectors and plastic scintillators, with a viewing angle of $\sim60\degree$ \citep{hassler2012}, and providing proton detections up to a stopping energy of 100 MeV. 
The original energy of the particle, $E$, is solved through analyzing $E$ versus $\mathrm{d}E/\mathrm{d}x$ correlations for each particle. Since RAD transmits the deposited energy in each triggered detector layer for almost all stopping protons, the particle identification is done in post-processing and is very accurate.
Protons stopping inside RAD can thus be selected and their intensities have been obtained in four energy channels: \mbox{12--24 MeV}, \mbox{24--40 MeV}, \mbox{40--60 MeV}, and \mbox{60--100 MeV}. 
The particles detected by RAD are a combination of primaries and secondaries resulting from spallation and energy losses as particles travel through the flight-time spacecraft shielding. The shielding distribution around RAD is very complex: most of the solid angle was lightly shielded with a column density smaller than 10 g/cm$^2$, while the rest was broadly distributed over a range of depths up to about 100 g/cm$^2$ \citep{zeitlin2013}. Due to this shielding, deducing the exact incident energies of particles as they reach the spacecraft is a challenging process. We briefly discuss correcting for these effects in section \ref{sec:comparisons2}.

Celestial mechanics dictate that a spacecraft on a Hohmann transfer to Mars remain magnetically well connected to Earth during most of its cruise phase \citep{posner2013}. This connection is also shown in Figure \ref{fig:heliospheric_ps}. Due to this reason, the intensity profiles seen at Earth and MSL are expected to show similar time evolutions.

\subsection{First arrival of particles and solar release time}

Intense energy release at the surface of the Sun or in the corona can accelerate SEPs to relativistic energies, allowing them to propagate rapidly along the Parker spiral \citep{Parker1958} to heliospheric observers. If the observer is magnetically well-connected to the acceleration site and particle transport is unhindered, the arrival time of first particles can be used to infer the travel distance, i.e., the Parker spiral length.

As each heliospheric location will see the first arrival of energetic protons at a different time, we have defined onset times separately for each spacecraft, listed in Table  \ref{tab:heliospheric_ps}. In Figure \ref{fig:flux}, the green vertical lines mark the onset times of the highest-energy channel corresponding to the arrivals of fastest protons. For STA and MES observations, we also define onset times of possible ESP events in low-energy channels, marked by grey lines, as will be discussed in more detail later. For STA, we find two distinct jumps, which may both be due to an ESP event. These times were defined from the raw data through subjective analysis of rise over a background level.

The nominal Parker spirals connecting the spacecraft to the Sun are shown in Figure \ref{fig:heliospheric_ps} assuming an average solar wind speed of \mbox{400 km s$^{-1}$} and their lengths have also been calculated and listed in Table \ref{tab:heliospheric_ps}. 
Given a Parker spiral length of \mbox{1.18 au} for an observer at Earth, 1 GeV protons (with a speed of $\sim 2.6 \times 10^5$ km/s) propagating from the flare site without scattering would arrive after \mbox{$\sim$670 s} or 11 minutes. A particle onset time at Earth at 01:56 would indicate a solar release time (SRT) of about 01:45 UT for these protons, which is consistent with radio burst observations \citep{Gopalswamy2013,Papaioannou2014}, considering the 8-min propagation time of radio signals from the Sun to \refbold{the} Earth. 
Table \ref{tab:heliospheric_ps} also lists the 1 GeV proton travel times and estimated associated SRTs, for each of the location considered, based on the calculated Parker spiral lengths. 
The observed MSL onset time is in good agreement with that at Earth and with the estimated proton release time, likely due to the good magnetic connection between the acceleration region and Earth/MSL. 
However, SRT values derived from MES, STA, and STB are very different from each other and hours later than the time of flare onset and shock formation. This indicates that these spacecraft were not magnetically well-connected to the solar acceleration site, and that particle transport to these locations was not due to propagation parallel to the magnetic field lines but was affected by drift motion, co-rotation, cross-field diffusion and turbulence effects. 

\subsection{Magnetic connectivity}\label{sec:connectivity}

The multi-spacecraft observations available for the SEP event on May 17th 2012 provide an exemplary chance to investigate magnetic connectivity between the Sun and observation platforms at a wide variety of longitudes and radial distances. We model magnetic connectivity by assuming the IMF to follow a Parker spiral. We use a constant solar wind speed of \mbox{400 km s$^{-1}$} for our modelling, based on the averaging described in Section \ref{sec:observations}.

\begin{figure*}[ht!]
\centering
\includegraphics[width=0.85\textwidth]{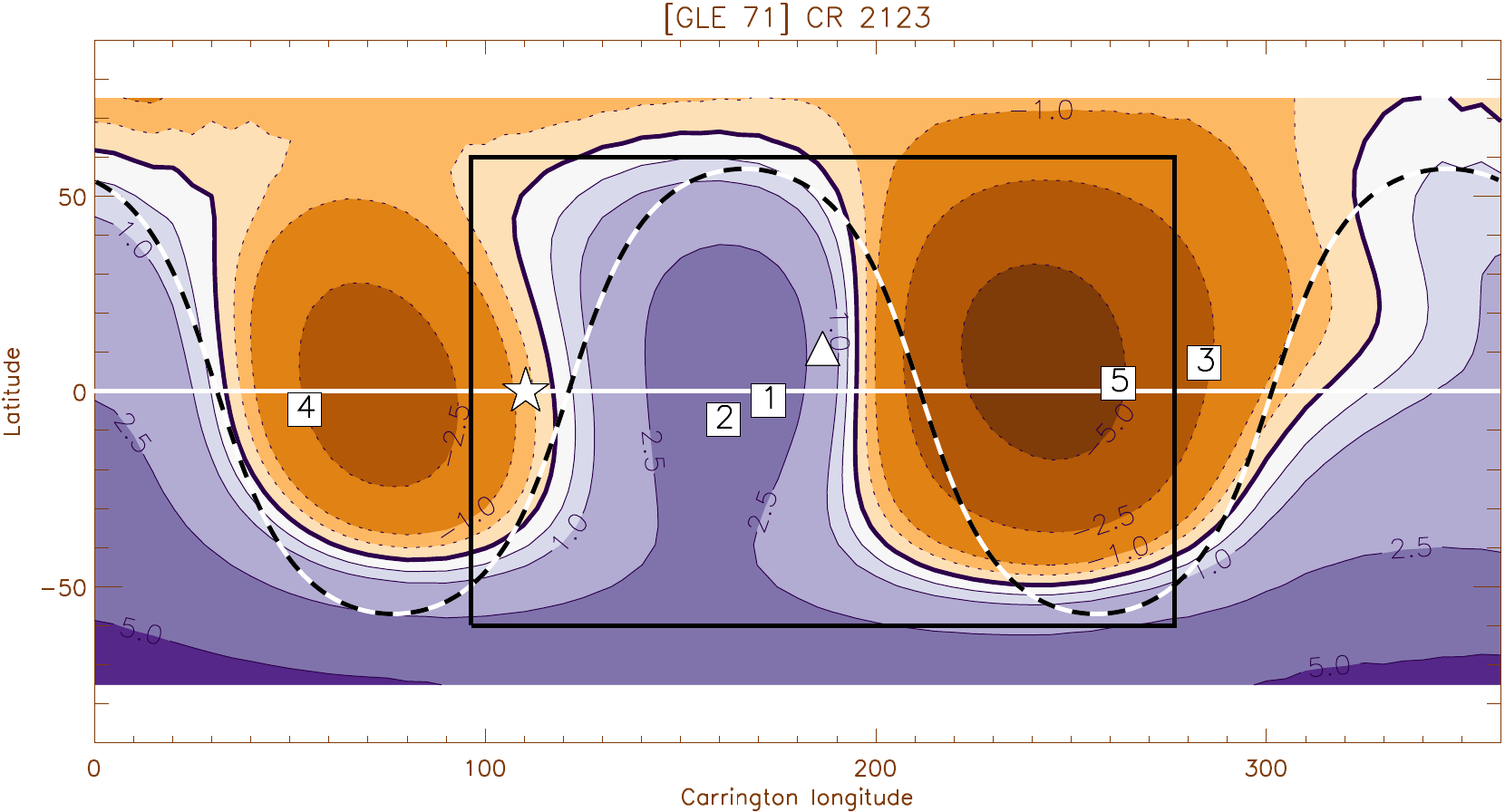}
\caption{Synoptic source surface map computed for $r=2.5\,R_\odot$ using photospheric measurements for Carrington rotation 2123. The location of the flare on May 17th, 2012, is indicated with a triangle. The central meridian at the time of the flare is indicated with a star. The Parker spiral connecting footpoints for each observer, assuming a solar wind speed of \mbox{400 km s$^{-1}$}, are shown with squares, numbered as \mbox{1: Earth}, \mbox{2: MSL}, \mbox{3: STA}, \mbox{4: STB}, \mbox{5: MES}. \refbold{Purple} regions indicate outward-pointing magnetic fields, \refbold{orange} regions inward-pointing magnetic fields, and the boundary line is shown as a solid \refbold{black} curve. \refbold{The heliographic equatorial plane is shown as a solid white line.} Contour values are given in microtesla. A fit for a simple wavy current sheet is shown as a black dashed curve, and the boundary of the injection region used in particle transport simulations is shown with a black rectangle. \refbold{Potential field} data is provided by the Wilcox Solar Observatory. }\label{fig:WSO_SYNOPTIC}
\end{figure*}

In Figure \ref{fig:WSO_SYNOPTIC}, we plot the Carrington Rotation 2123 solar synoptic source surface map \citep{Hoeksema1983} for $r=2.5\,R_\odot$, resulting from potential field modelling, provided by the Wilcox Solar Observatory. The model assumes a radial magnetic field at the solar surface and at $r=2.5\,R_\odot$. The plot shows the location of the flare on May 17th 2012 (indicated by a triangle) relative to the central meridian, along with estimated Parker spiral footpoints for the five observation platforms. As the plot shows, Earth (labelled 1) and MSL (2) are connected to regions on the Sun's surface very close to each other, with STA (3) and MES (5) connected to more western longitudes, close to each other. STB (4) is connected to more eastern longitudes.

Figure \ref{fig:WSO_SYNOPTIC} also includes, as a thick white solid curve, a depiction of \refbold{a potential field model} neutral line between hemispheres of outward and inward pointing magnetic field. A model of a simple parametrized wavy neutral line\refbold{, based on a tilted dipole formulation,} is fitted to this \refbold{neutral} line using a least squares \refbold{distance} fit method, as described in \cite{Battarbee2018}. This neutral line parametrisation is the $r=2.5\,R_\odot$ anchor point for our model wavy HCS, and the wavy HCS parameters are described in section \ref{sec:simulations}. Finally, figure \ref{fig:WSO_SYNOPTIC} shows a rectangular region of width $180^\circ$, extending to latitudes $\pm 60^\circ$, which we use as a model injection region for SEPs. The width of the injection region was iterated upon, until an agreement between observations and simulations, for as many heliospheric observers as possible, was achieved.

As the solar wind flows outward and the solar surface rotates, magnetic structures at a given heliocentric distance are co-rotated westward. In Figure \ref{fig:WSO_SYNOPTIC}, this would be described by the \refbold{potential field} polarity map including the HCS moving to the right. We validate the synoptic source and Parker spiral model through simple radial magnetic field observations. MES and STA are in regions of inward-pointing magnetic field throughout the analysed time period, in agreement with the map. Up until the time of the flare, Earth is connected to outward-pointing field lines, after which a strong interplanetary CME (ICME) is detected and the field orientation flips. STB is initially connected to inward-pointing field lines, but from the 19th of May onward, the direction points inward, in agreement with the spacecraft crossing the HCS.

\subsection{Interplanetary shocks and energetic storm particles} \label{sec:IPshock}

In addition to SEPs accelerated close to the Sun during the initial, strong phase of the solar eruption, particle acceleration can happen throughout the inner heliosphere at propagating interplanetary (IP) shocks, driven by ICME fronts. Depending on the heliospheric location relative to the flare site and the ICME, different spacecraft see different properties of the event. The time profiles of \textit{in-situ} measurements in Figure \ref{fig:flux} and estimated SRTs in Table \ref{tab:heliospheric_ps} suggest that the particle intensities at Earth and MSL (with estimated SRTs of 01:45 and 01:46) are dominated by coronally accelerated SEPs, but at MES and STA, there is an additional population of energetic storm particles (ESPs) accelerated by an IP shock. To identify and decouple the signal of particles accelerated at an IP shock from those accelerated early on in the corona, we turn to ICME and shock catalogues. 


For MES, the circum-Mercurial orbital period of only 8 hours and related magnetospheric disturbances make identification of ICMEs challenging. \cite{Winslow2015} were able to detect an ICME at MES, lasting from 12:09 until 15:38 on May 17th. The shock transit speed was identified as \mbox{1344 km s$^{-1}$}. 
ESPs usually peak at lower energies than coronally accelerated SEPs, and are found only in the vicinity of the IP shocks due to turbulent trapping. At MES shown in Figure \ref{fig:flux}, we notice a clear intensity peak, likely due to ESPs, starting around 12:10 marked by a grey line right after the arrival of the ICME. 


A comprehensive catalogue of ICMEs, IP shocks, and streaming interactive regions (SIRs) for the STEREO spacecraft has been compiled by \cite{Jian2013}\footnote{\url{http://www-ssc.igpp.ucla.edu/forms/stereo/stereo_level_3.html}}. A shock was detected at STA on May 18th at 12:43, followed by an ICME until 09:12 on May 19th. The deka-MeV proton channels at STA show major enhancements starting at about 15:25 on May 18th (marked by a grey line), which can be attributed to IP shock accelerated ESPs. A smaller enhancement is seen at 04:52, possibly due to a foreshock of ESPs escaping in front of the IP shock.

STB is reported to be within a SIR from 23:48 on May 18th until 16:35 on May 22nd, well after the weak increase in proton flux. Upon further inspection of relevant solar wind measurements at STB, the possibility of an IP shock passing the spacecraft on between the 18th and 19th of May cannot be ruled out, but the data are ambiguous. 
An alternative explanation for the particle enhancement at STB, which begins less than 12 hours after the flare, is for coronally accelerated particles to drift there along the HCS, which co-rotates over the position of STEREO-B. We include this HCS drift in our simulations and assess this possibility in section \ref{sec:comparisons1}.


Many spacecraft are available for observing near-Earth transients. Both Wind and ACE databases report the Earth as within an ICME already from the 16th of May, being thus unrelated to the GLE 71 eruption. The Wind ICME list\footnote{\url{https://wind.nasa.gov/2012.php}} lists the ICME starting at May 16th 12:28, and ending at May 18th 02:11. ACE observations by \cite{Richardson2010}\footnote{\url{http://www.srl.caltech.edu/ACE/ASC/DATA/level3/icmetable2.htm}} list an ICME starting on May 16th 16:00, and ending at May 17th 22:00 UT. The only assertion of an actual shock is from the ACE list of disturbances and transients (see \citealt{McComas1998} and \citealt{Smith1998})\footnote{\url{http://espg.sr.unh.edu/mag/ace/ACElists/obs_list.html}}, with a shock at May 17th 22:00, but it is registered only in magnetic field measurements. Thus, we find no suggestion that there should be a significant ESP signal at Earth.

At the location of MSL, neither magnetic nor plasma measurements are available to identify ICMEs and IP shocks. No ESP structures are present in the RAD data. However, as RAD measurements of low energy protons are affected by nonuniform shielding, we cannot rule out the possibility of an ICME associated shock passing at the location of MSL. 

\section{Particle transport simulations}\label{sec:simulations}

In order to model heliospheric transport of SEPs accelerated during the May 17th event, we simulated the propagation of $3\cdot10^6$ test particle protons, from the corona into interplanetary space, using the full-orbit propagation approach of \cite{Marsh2013} and \cite{Marsh2015}. This model naturally accounts for particle drifts and deceleration effects, and allows for the generation of virtual time profiles at many heliospheric observer locations. Our model was newly improved by the inclusion of a HCS, normalised to a thickness of \mbox{5000 km} at \mbox{1 au}, as introduced in \cite{Battarbee2017} and as extended to non-planar geometries in \cite{Battarbee2018}. \refbold{We present here the first results of three-dimensional forward modelling of SEP propagation, extending throughout the inner heliosphere, for this event. Because we focused on multi-spacecraft observations and the 3D spatial distribution of particle fluxes, we have not performed comparisons with 1D modelling efforts of large SEP events (see, e.g., \citealt{Kocharov2017})} 


We inject energetic particles into our transport simulation assuming acceleration to happen at a coronal shock-like structure. Acceleration efficiency across a coronal shock front is a complex question in its own right, with applications to the event in question presented in \cite{Rouillard2016} and \cite{Afanasiev2017}. Their analysis of CME expansion suggests a CME width of 100 degrees in longitude with varying efficiency along the front. Using this width, our simulations had difficulty recreating proton time profiles at many heliospheric observer locations. Thus, we chose to assume additional spread of energetic particles in the corona during the early phase of the event. We iterated the width of the injection area, finding one of $180^\circ$ width in longitude, centered at the flare location, to provide the best results \refbold{when attempting to recreate observed time profiles. This wide injection region is in agreement with a very wide coronal shock acting as the source of accelerated particles}. Injection was performed between equatorial latitudes of $\pm 60^\circ$. The injection region is shown in Figure \ref{fig:WSO_SYNOPTIC} as a black rectangle.

In order to decouple injection and transport effects, we chose to model particle injection through a simplified case. Thus, we inject isotropic protons from the aforementioned region with a uniform source function at a heliospheric height of \mbox{$r=2.0\,R_\odot$}, at the estimated solar particle release time of 1:40 \citep{Gopalswamy2013}. As most acceleration of particles is assumed to take place at low heliospheric heights of up to a few $R_\odot$, an instantaneous injection is a fair approximation. Any ESPs accelerated by the interplanetary shock are not modelled. Protons were injected according to a power law of \mbox{$\gamma=-2.0$}, distributed in the energy range \mbox{$10-600$ MeV}. \refbold{The chosen power law is close to the value derived by \citep{Kuhl2016} from in-situ observations using SOHO/EPHIN. As our focus was on multi-spacecraft observations and modelling over a large spatial extent, }we did not model protons in the GeV energy range due to \refbold{comparison data from GeV-range observations being available only in the vicinity of the Earth. Extending our injection power law higher, whilst maintaining adequate statistics, would have required computational resources beyond the scope of this project}. The total simulation duration was set to 72 hours.

During transport, we modelled particle scattering using Poisson-distributed scattering intervals, with a mean scattering time in agreement with a mean free path of \mbox{$\lambda_\mathrm{mfp}=0.3\,\mathrm{au}$}. Particles experience large-angle scattering in the solar wind frame for which we used a constant radial solar wind speed of \mbox{$u_\mathrm{sw}=400\,\mathrm{km}\,\mathrm{s}^{-1}$} throughout. The magnetic field was scaled to \mbox{$B=3.85\,\mathrm{nt}$ at 1 au}, consistent with observations. For the winding of the magnetic field, we assumed an average solar rotation rate of \mbox{$\Omega_\odot = 2.87\times10^{-6}\,\mathrm{rad}\,\mathrm{s}^{-1}$} or 25.34 days per rotation. 

In order to model particle detection at spacecraft, we gathered simulated particle crossings across virtual observer apertures at the locations of STA, MES, Earth, MSL, and STB. For each virtual observer, we used energy bins in agreement with those listed in section \ref{sec:observations} and time binning of 60 minutes. To increase statistics, simulated protons propagating outward from the Sun were gathered over a $10^\circ\times 10^\circ$ angular window at each observer location. As the orbital period of Mercury is only 88 days, we implemented longitudinal orbital motion of virtual observers around the Sun. Due to the large time bins used, we have not attempted to infer exact onset times from particle simulations, \refbold{nor have we explicitly considered twin acceleration scenarios (see, e.g., \citealt{Ding2016} and \citealt{Shen2013})}.

For parametrization of the wavy current sheet, we used a least squares sum method to fit the distance of the $r=2.5\,R_\odot$ \refbold{potential field} neutral line to a wavy model neutral line, resulting in \refbold{a dipole tilt angle} of $\alpha_\mathrm{nl}=57^\circ$, \refbold{a longitudinal} offset of $\phi_\mathrm{nl}=101^\circ$, and an \refbold{peak count} multiplier of $n_\mathrm{nl}=2$. This source neutral line at $r=2.5\,R_\odot$, used as the anchor point of the current sheet, is depicted in Figure \ref{fig:WSO_SYNOPTIC} as a dashed black curve. 

\begin{figure}[ht!]
\centering
\includegraphics[trim=60 150 0 150,clip, width=0.43\textwidth]{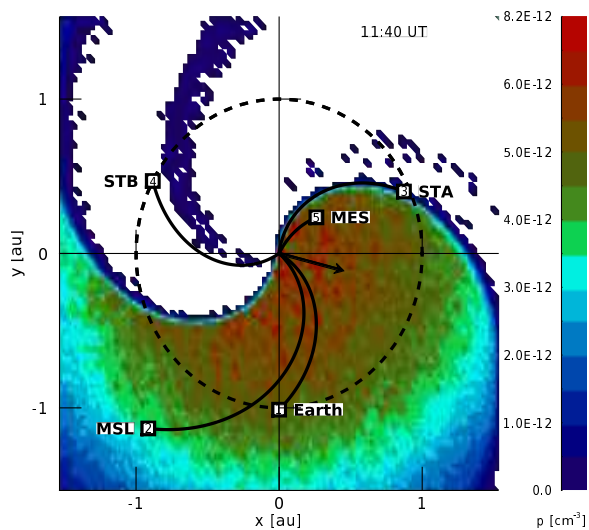}
\caption{Filled contours of simulated particle density in units cm$^{-3}$ in the inner heliosphere, close to the ecliptic, 10 hours after injection (11:40 UT). The locations of five observer platforms are shown along with Parker spiral connectivity using a solar wind speed of \mbox{400 km s$^{-1}$}. The \mbox{1 au} distance is shown with a dashed circle. The arrow is placed along the radial direction at the flare location.}\label{fig:SPEC_xy}
\end{figure}

Figure \ref{fig:SPEC_xy} shows the ecliptic distribution of accelerated protons, 10 hours after injection (11:40 UT), along with observer locations and Parker spiral connectivity assuming a solar wind speed of \mbox{400 km s$^{-1}$}. Shaded contours show the scaled particle density in units $\mathrm{cm}^{-3}$ between $-20$ and $+20$ degrees latitude. Of particular interest is the band of protons close to STB, which have experienced HCS drift.

As our simulations do not include a background intensity and provided counts in arbitrary units, the particle densities and intensities had to be calibrated using a normalisation multiplier. Due to good magnetic connectivity at Earth, we decided to use a near-Earth peak intensity as the reference intensity. For this normalisation, \refbold{we used the \mbox{38.0--82.0 MeV} GOES energy channel, as it had a clearly defined peak. Although the background levels at GOES were enhanced due to magnetospheric effects, we assume that the peak values were not affected significantly.}
Hereafter, for all time profile and peak intensity analysis, results from our simulations were multiplied by a single normalisation constant, which resulted in agreement between peak intensities deduced from the \refbold{\mbox{38.0--82.0 MeV}} channels at Earth from both simulations and observations.

\subsection{Comparison with observations: time profiles}\label{sec:comparisons1}

\begin{figure*}[ht!]
\centering
\includegraphics[trim=50 30 50 30,clip, width=\textwidth]{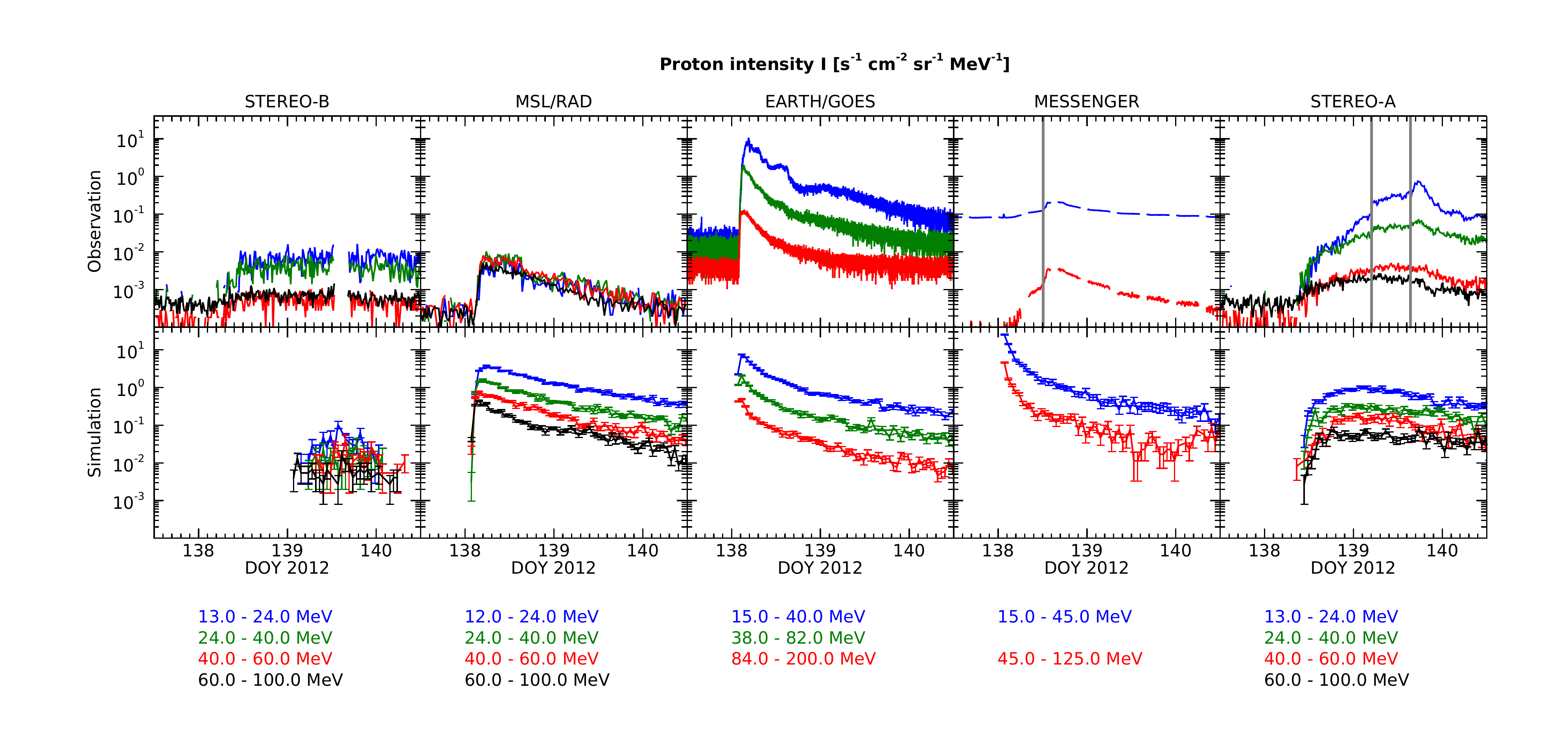}
\caption{Top row: proton time profiles at five heliospheric locations. Bottom row: corresponding virtual time profiles generated through SEP transport simulations. The locations are ordered according to connected footpoint longitude: STB, MSL, Earth, MES, and STA. MES and STA observations are also marked with the onsets of ESP-related proton effects with grey lines. For STA, the first grey line designates the estimated onset of foreshock ESP flux.}\label{fig:SPEC_COMBINED}
\end{figure*}

In this section, we compare the intensity time profiles of simulations and observations. Figure \ref{fig:SPEC_COMBINED} displays results of both observations and simulations, with intensity time profiles for selected energy bins at each location, actual observations on the top row and simulation results on the bottom row. \refbold{For simulation time profiles, we include error bars calculating an estimate of uncertainty for the intensity using the square root of registered particle counts.} Panels are ordered according to observer footpoint longitude, as shown in Figure \ref{fig:WSO_SYNOPTIC}. 
We first focus on the \refbold{qualitative} shape of the time profiles, proceeding from west to east (right to left) in observer footpoint longitude.

At STA, observations show a gradually increasing flux, and SRTs calculated from onset times in Table \ref{tab:heliospheric_ps} are many hours after the flare time. This suggests that the location of STA does not have good magnetic connectivity to the injection region \refbold{at the start of the event}. 
However, the numerical simulation is able to provide a proton time profile in agreement with observations, 
using the 01:40 UT release time. Protons fill the well-connected field lines with a population which isotropizes, and this population is then co-rotated over the STA position\refbold{, becoming magnetically well-connected later in the simulation}. STA observations in the lowest two energy bins show an additional feature, with bumps in intensity at approximately 04:52 and 15:25 on DOY 139. Both these bumps are designated with grey vertical lines in Figure \ref{fig:SPEC_COMBINED}. As described in section \ref{sec:IPshock}, an IP shock is detected at STA, and these enhancement at low energies can be explained as ESPs related to a passing IP shock. The first bump would indicate the arrival of an enhanced foreshock region, and the second bump would occur during the actual shock crossing. The simulated results do not show these bumps as ESP \refbold{enhancements} were not modelled by the SEP transport simulations. 

\refbold{At MES, the rapid increase in particle intensity of our simulations does not agree with the observed delayed particle flux}. The simulated time profile shows a simple abrupt event due to an efficient connection to the injection region, although it does drop off fast as the observer is rotated westward around the Sun with a rapid 88 day orbital period. Observations seem to suggest that coronally accelerated particles were not propagated efficiently to MES, as the enhancement over background intensities is small and happens too late. Shielding effects due to Mercury or its magnetic field were accounted for by masking out measurements with altitudes below \mbox{5000 km}. Thus, if an abrupt coronally accelerated component had been present at the position of MES, we should have \refbold{detected} it. A delayed enhancement, possibly due to ESPs, has a good match with the reported ICME crossing at 12:09, preceded by a foot of particles accelerated at the IP shock. This enhancement appears stronger in the \mbox{45--125 MeV} channel, which might indicate that the signal at MES is strongly influenced by particle drifts, as the magnitude of particle drifts scales with energy. Alternatively, the signal in the \mbox{15--45 MeV} channel might be hidden behind a strong background contamination signal. We note again that ESPs were not modelled in our transport simulations. \refbold{We also note that although we only show derived \mbox{15--45 MeV} and \mbox{45--125 MeV} energy channels in figures \ref{fig:flux} and \ref{fig:SPEC_COMBINED}, the single coincidence channel for \mbox{$> 15$ MeV} detection did not show an abrupt rise, but rather a similar time profile as the shown derived energy channels. This also rules out single coincidence channel contamination as a source of discrepancy.}



\refbold{After discussing the observed and simulated time profiles at STA and MES, it is appropriate to recall the assumed magnetic connectivity to these observers based on Figures \ref{fig:heliospheric_ps} and \ref{fig:WSO_SYNOPTIC}. The magnetic connectivity footpoint of MES is eastward of the STA connectivity footpoint, i.e. closer to the flare location. Thus, assuming a Parker spiral IMF and a simplistic injection region surrounding the flare location, a strong particle signal at STA should suggest a strong signal also at MES. This is in agreement with our simulated results, but in clear disagreement with the observations.}

\refbold{One possible explanation for the discrepancies between observations and simulations is that the IMF shape may differ from that of a Parker spiral. We find that STA was in a fast solar wind stream prior to the event, and additionally a SIR was detected at STA on May 16th \citep{Jian2013}, with a maximum solar wind speed of \mbox{$660 \, \mathrm{km}\,\mathrm{s}^{-1}$}, well above the value of \mbox{$400 \, \mathrm{km}\,\mathrm{s}^{-1}$} used in our simulations. Thus, the IMF might have been primed by this stream, providing STA with a connected footpoint significantly east of the one used in our model. As we do not have solar wind speed measurements at Mercury, we cannot make similar educated guesses about the longitudinal position of the well-connected footpoint location for MES.}

\refbold{Another possible explanation is that smaller-scale effects of the IMF and particle propagation are taking place, invalidating the Parker spiral model. Recent research into field-line meandering (see, e.g., \citealt{Laitinen2016} and \citealt{Ruffolo2012}) and SEP cross-field propagation (see, e.g., \citealt{Laitinen2013} and references therein) has investigated this problem. New missions going close to the Sun will provide key data to validate these theories. Recent research, shown in in panel (a) of Figure 6 in \cite{Laitinen2017}, suggests, however, that the early-time cross-field variance of a particle distribution is strongly dependent on radial distance. Thus, if we assume a narrowed injection region, during the early phase of the event, STA could be connected to the injection region through widely meandering field lines, whereas MES at a distance of only \mbox{$0.34\,R_\odot$} would remain outside this region.}

At the location of Earth, we compare \refbold{three GOES} 
energy channel time profiles with observations. 
The highest energy channel at \refbold{\mbox{84.0--200.0 MeV}} provides an excellent match between simulations and observations, suggesting acceleration was near-instantaneous in the corona, and that Earth was well-connected to the acceleration region. At the \refbold{middle energy channel of \mbox{38.0--82.0 MeV}, the agreement between simulations and observations is also good, although the observed time profile begins to decrease slightly more rapidly than the simulated one. This may be due to, e.g., differences in particle scattering rates early in the event. At the lowest energy channel of \mbox{15.0--40.0 MeV}, agreement is moderately good, although the rate of intensity decay is slightly different for observations and the simulation. Additionally, an enhancement in observed intensity is found about halfway through DOY 138. Although databases of interplanetary shocks showed only weak indications of a shock passing at earth, an IP-shock related ESP event is still the most likely explanation for this feature.}

At MSL, with a similar magnetic connection to Earth, time profiles agree \refbold{moderately} well with simulations. The observations at MSL seem to show similar intensities for all the different channels, resulting in a near-flat spectrum. The total intensities observed at the detector are more than an order of magnitude lower than the simulated intensities. However, as the general shape of the time profile agrees well with that of simulations, we suggest that transport and connectivity is not the primary cause of the disrepancy, but rather, that is due to the flight-time spacecraft shielding around MSL/RAD, causing particles to decelerate, fragment, or be deflected away. 
Modelling this effect in detail and performing inversion on the measured particle flux is rather challenging. We present preliminary corrections accounting for the energy loss of protons in section \ref{sec:comparisons2}. 


Although the footpoint of STB is separated from the flare region by almost 180 degrees, a weak enhancement in proton flux is seen both in observations and in simulation results. There was a SIR in the vicinity of STB in the time period following the event \citep{Jian2013}. Due to this and complicated solar wind observations, a weak ICME-driven shock cannot be ruled out. However, the most likely candidate for explaining the SEP flux enhancements at STB is coronally accelerated particles transported along the HCS. The successful simulation of this signal at STB is only possible through the results of our newly improved SEP transport simulation, supporting an IMF with two magnetic polarities separated by a wavy HCS. Particles propagate along the HCS, which is co-rotated over the position of STB (see Figure \ref{fig:WSO_SYNOPTIC}). The difference in onset time and signal duration between simulations and observations can be explained by inaccuracies in the exact position and tilt of the HCS at the position of STB. 

\subsection{Comparison with observations: peak intensities}\label{sec:comparisons2}

\begin{figure*}[ht!]
\centering
\includegraphics[trim=0 20 0 0,clip, width=0.75\textwidth]{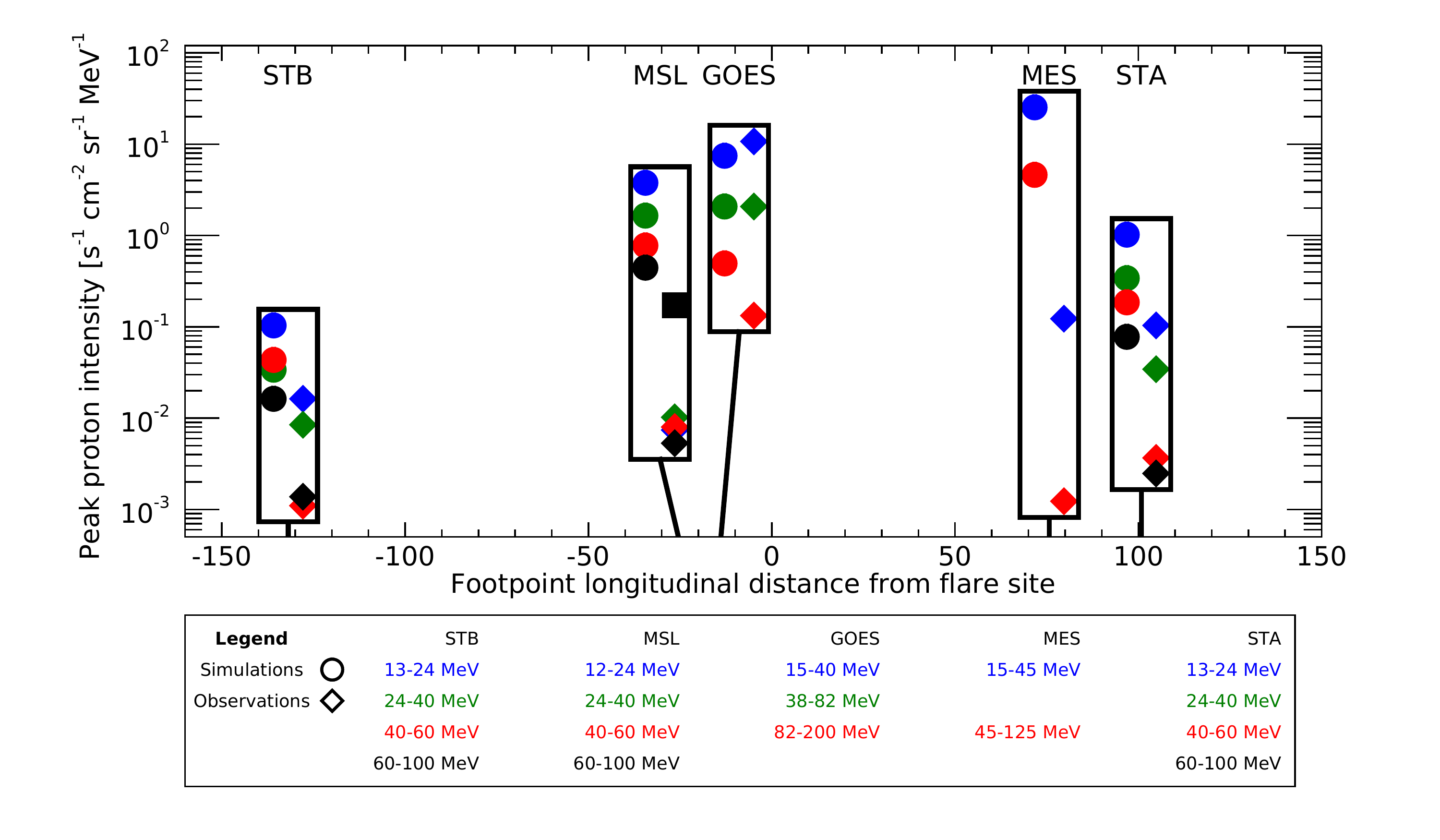}
\caption{Peak intensities of time profiles, recorded from observations (circles) and numerical simulations (diamonds). Observers are placed on the x-axis according to footpoint longitudinal distance from the flare. Peak intensities inferred from observations excluded time periods with ESP effects. 
For MSL/RAD, in addition to the recorded intensities, we show as a black square a version of the \mbox{60-100 MeV} proton channel, with preliminary first-order corrections for energy losses due to spacecraft shielding.}\label{fig:peakfluxcomparison}
\end{figure*}

In order to further assess longitudinal accuracy of our SEP transport simulations, we gathered peak intensities for both simulations and observations for each channel and plotted them according to estimated footpoint location (see also Figure \ref{fig:WSO_SYNOPTIC}). The peak intensities for STB, MSL, \refbold{GOES}, MES, and STA are shown in Figure \ref{fig:peakfluxcomparison}, along with peak intensities deduced from simulations. In determining observational peak intensities for STA and MES, we excluded time periods deemed to be enhanced by ESP effects. For STA, this exclusion \refbold{began at} \mbox{04:52 UT} on the \mbox{18th of May}, corresponding with the foreshock region of the IP shock. This foreshock region is visible especially in the \mbox{13--24 MeV} channel, but somewhat also in the \mbox{24--40 MeV} channel.

Comparing the \refbold{\mbox{15--40 MeV} and \mbox{38--82, MeV}} observed and simulated peak intensities at Earth results in a good match due to \refbold{the} \refbold{\mbox{38--82 MeV}} energy channel being used for the normalisation of simulation results. However, observations at \refbold{\mbox{82--200 MeV}} show smaller intensities than the respective simulation results. \refbold{We discuss the effects of the injection spectrum on peak intensities at the end of this section.}

At STB, \refbold{simulated peak intensities exceed observed intensites by approximately one order of magnitude, but all channels show a similar intensity offset}. All channels at STB show only a weak increase over background intensities, which is modelled well by the HCS-transported particles in the simulation. \refbold{The highest two observed energy channels are somewhat lower than the simulated ones, suggesting an injection spectrum related effect, similar to what was seen at Earth.}

At STA, after excluding all ESP-enhanced regions from observations, observed and simulated peak intensities \refbold{show a similar order-of-magnitude difference as was noted at STB}. Similar to STB, the observations in the two highest energy channels exhibit slightly weaker peak intensities\refbold{, pointing to the injection spectrum as the culprit.} 

Neither the time profiles nor the peak fluxes of simulations and observations at MES agree with each other, which indicates that the true magnetic connectivity to MES is more complicated than the one used in our simulations. Based on our calibrations, we believe \refbold{that instrumental effects cannot explain this discrepancy}. The simulated injection region was set to a width of $180^\circ$ in order to provide a good time profile match at STA, however, CME modelling from observations produced shocks fronts of only $100^\circ$ width. A narrower injection region might prevent coronally accelerated particles from reaching MES\refbold{, but would also result in a poor match for STA}. The question of magnetic connectivity from the corona to STA and MES was explained in detail in section \ref{sec:comparisons1}. If the CME \refbold{were to transition} to an ICME, and further \refbold{from the Sun}, expand in width, this could be seen as ESPs at MES, thus explaining the observations.

At MSL, the observed peak intensities are much lower than those of simulations, \refbold{likely} due to the in-flight shielding covering much of the detector. As a first step toward correcting particle fluxes at MSL/RAD, we performed calculations of the energy loss of protons traversing a model of the spacecraft shielding. Proton energy losses in matter are primarily due to ionization, which is characterized by the Bethe-Bloch equation, which was used in our calculations. We considered the distribution of aluminium equivalent shielding depth within RAD's viewing angle \citep{zeitlin2013}. Due to the involved complexity, we did not account for generation of secondary particles, which play a major role at low energies. Thus, we produced a corrected peak intensity only for the \mbox{60-100 MeV} channel, shown in Figure \ref{fig:peakfluxcomparison} as a black square. This value appears to be a \refbold{better} match with both simulation results and \refbold{GOES} observations\refbold{, showing a similar relationship to the simulated channel as was seen for the GOES \mbox{82--200 MeV} channel.} Recreating original particle intensities at all channels of MSL/RAD will be the topic of future investigations.

\refbold{In comparing peak intensities for observations and simulations, many things must be taken into account. At MSL, shielding weakens the observed intensity in a significant manner, which requires post-processing to correct for. Magnetic connectivity at MES provides contradicting time profiles and peak intensities. At STA and STB we are able to reproduce time profile shapes, but the peak intensities are over-estimated in our transport simulations. However, noting that our injection source was uniform in longitude, which is not a realistic estimate, but allows us to now draw conclusions from the peak intensity fits. At longitudes close to the flare location, injection was as simulated and normalised, but at longitudes far away from the flare, injection efficiency drops, apparently an order of magnitude. This would be unsurprising, considering our injection region was set at 180 degrees. Thus, we now have indication that a strong injection takes place at the observed shock front with a width of ca. 100 degrees, but early-time propagation effects spread particles to a region of up to 180 degrees with lesser intensity.}

\refbold{A general trend was that simulated and observed fluxes for low energy channels were in better agreement than those of higher energy channels. This suggests that our simulated injection power law of $\gamma=2$ was too hard, and the actual solar eruption had injected fewer high-energy particles than simulated. From our fitting, we can deduce that either a softer injection spectrum or a broken power law with weaker injection at high energies is likely to be closer to the truth.}

\refbold{Overall, the simulated peak intensities presented in Figure \ref{fig:peakfluxcomparison} show that the 3-dimensional propagation simulations have great merit in increasing our understanding of large SEP events. By correctly accounting for particle drifts, we can simulate propagation of particles over a wide range of energies, and thus, make educated estimates regarding the injection power law at the Sun. The general good agreement between how peak intensities are grouped according to footpoint location suggests that both prompt (such as Earth and MSL) and delayed (such as STA) SEP fluences can be modelled, once a longitudinal injection efficiency dependence is found. For this purpose, work such as that done by \cite{Rouillard2016} and \cite{Afanasiev2017} is very useful. The mismatch between observations and simulations at Mercury/MESSENGER shows that the inner heliosphere is a complicated environment and proper modelling of magnetic connectivity throughout it requires additional effort.}

\subsection{Comparison with observations: pitch-angles}\label{sec:comparisons3}

\begin{figure}[htb!]
{\includegraphics[trim=0 20 0 0,clip, width=0.45\textwidth]{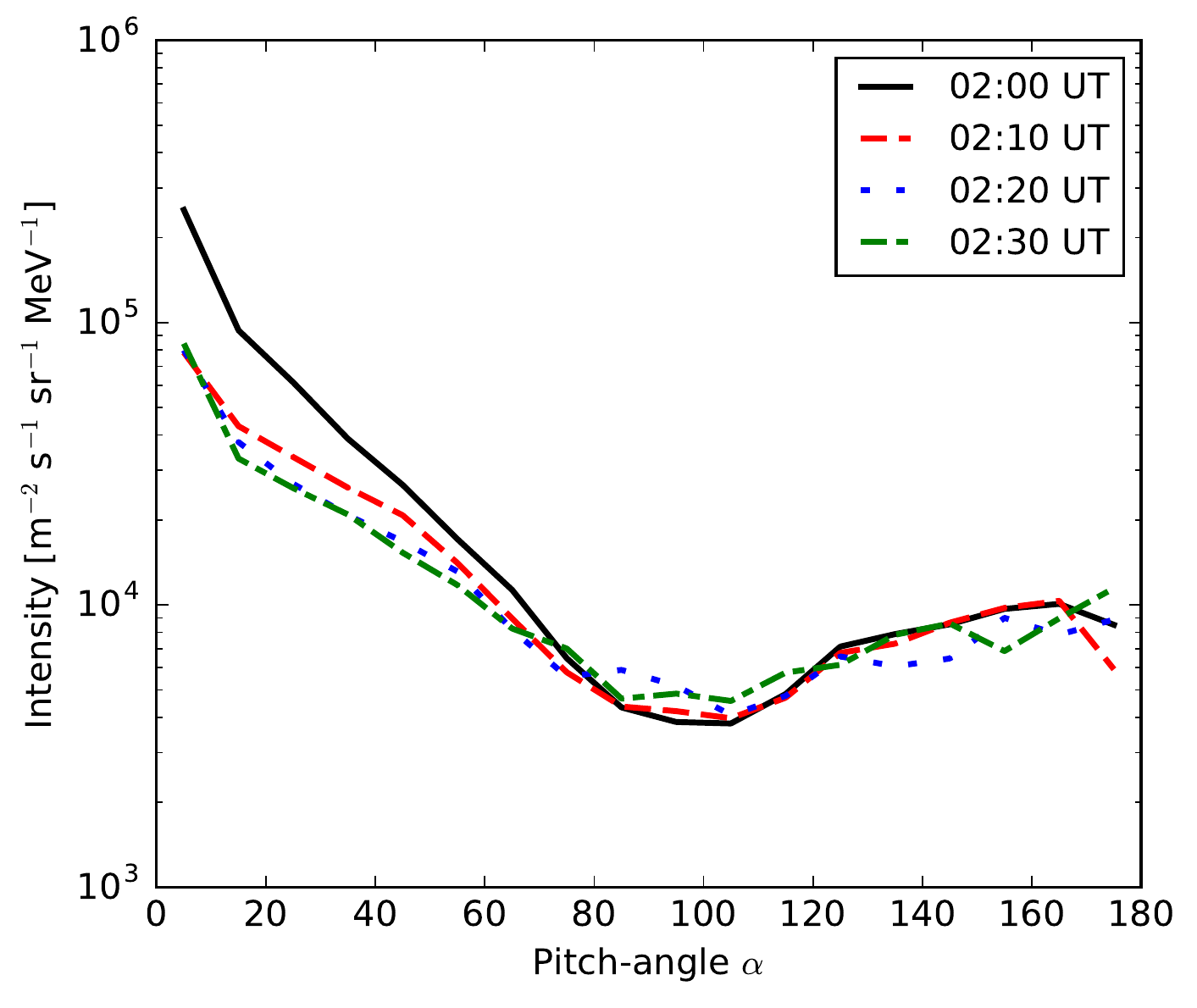}}
\caption{Derived pitch-angle distribution for \mbox{10--600 MeV} protons at Earth, using particle propagation simulation data for the early phase of GLE 71. Time in UT refers to the start of each 10 minute binning interval.}\label{fig:pitchangles}
\end{figure}

Observations of the pitch angle distribution of GeV-class particles for GLE 71 have shown an unusual twin-beam distribution (\citealt{adriani2015pamela}, \citealt{Mishev2014}). In our model, we have used a simplified model of the scattering experienced by the SEPs, by considering only large angle scattering events. In figure \ref{fig:pitchangles} we show the derived \mbox{10-600 MeV} proton pitch-angle distribution at Earth for the early phase of the simulation. We gathered proton crossings across the \mbox{1 au} sphere at the location of Earth, gathering crossings over a $10^\circ\times 10^\circ$ angular window and applied the same intensity scaling as for earlier plots. We also performed scaling to account for solid angle size for each bin. The results indicate that our model is capable of reproducing a twin-stream distribution without including additional magnetic structures such as loops associated with preceding CMEs. We note that some qualitative similarities with figure 6 of \cite{Mishev2014} exist, but more detailed analysis would require refining our scattering model.

\section{Conclusions}\label{sec:conclusions}

We have presented extensive, detailed multi-spacecraft observations of proton intensities for \refbold{the solar eruption of} May 17th, 2012. We have shown the event to encompass a large portion of the inner heliosphere, extending to a wide range of longitudes, with a strong detection at Earth, MSL, and STA. We were able to analyse SEP transport and magnetic connectivity based on a new improved 3D test particle model.

Our SEP transport model solves the full-orbit 3D motion of test particle SEPs within heliospheric electric and magnetic fields. The model naturally accounts for co-rotation, particle drifts and deceleration effects (\citealt{Marsh2013}, \citealt{Dalla2013}, and \citealt{Dalla2015}). Our new improved model includes, for the first time, effects due to a solar magnetic field of two different polarities, separated by a wavy HCS. We model proton injection with a shock-like structure near the Sun, and model interplanetary transport in accordance with a particle mean free path of $\lambda_\mathrm{mfp}=0.3\,\mathrm{au}$.


We present novel multi-spacecraft analysis of an SEP event, encompassing all heliolongitudes and radial distances ranging from \mbox{0.35 au} to \mbox{1.46 au}. We compare results from multiple spacecraft and particle detectors with virtual observers placed within a large-scale numerical simulation. We improve upon previous studies, usually focused on a single observation platform, with our analysis, providing good agreement between simulations and observations at multiple heliospheric locations.

We show that for GLE 71, observers magnetically connected to regions close to the flare location exhibit a rapid rise in proton intensity, followed by a prolonged fall-off. We report how STEREO-A observations are explained through a combination of co-rotation of an SEP-filled flux tube across the spacecraft in combination with an ESP event, and how STEREO-B observations can be explained through HCS drift of coronally accelerated protons.

For four out of five observer locations, we are able to find a good match in both the qualitative intensity time profiles and the quantitative peak intensities when comparing observations and numerical simulations\refbold{, if we assume that injection efficiency weakens as a function of longitudinal distance from the flare location}. Our results suggest modern modelling of large-scale \refbold{solar eruptions} has improved, and has benefited greatly from the opportunities provided by the two STEREO spacecraft, as well as other heliospheric and even planetary missions such as MESSENGER and MSL. SEP forecast tools such as those presented in \cite{Marsh2015} should play an important role in furthering our understanding of solar activity.

Our study shows that magnetic connectivity to the injection region as well as the perpendicular propagation of particles in interplanetary space are important factors when assessing the risk of SEP events. Solar wind streams, interacting regions, and concurrent coronal mass ejections with associated magnetic structures alter the IMF and particle transport conditions, yet modern computation methods are capable of impressive modelling of SEP events. Further improvements in modelling of the background conditions for SEP simulations are required, with 3D magnetohydrodynamic models a likely candidate for future studies. 

\begin{acknowledgements}
This work has received funding from the UK Science and Technology Facilities Council (STFC; grant ST/M00760X/1) and the Leverhulme Trust (grant RPG-2015-094). 
We acknowledge the International Space Science Institute, which made part of the collaborations in this paper through the ISSI International Team 353 "Radiation Interactions at Planetary Bodies". 
RAD is supported by the National Aeronautics and Space Administration (NASA, HEOMD) under Jet Propulsion Laboratory (JPL) subcontract 1273039 to Southwest Research Institute and in Germany by DLR and DLR’s Space Administration grant numbers 50QM0501, 50QM1201, and 50QM1701 to the Christian Albrechts University, Kiel. The RAD data used in this paper are archived in the NASA Planetary Data Systems Planetary Plasma Interactions Node at the University of California, Los Angeles. The PPI node is hosted at http://ppi.pds.nasa.gov/.
The Solar and Heliospheric Observatory (SOHO) is a mission of international collaboration between ESA and NASA. Data from the SOHO/ERNE (Energetic and Relativistic Nuclei and Electron) instrument was provided by the Space Research Laboratory at the University of Turku, Finland. 
Wilcox Solar Observatory data used in this study was obtained via the web site \url{http://wso.stanford.edu} at 2017:01:12\_05:40:05 PST courtesy of J.T. Hoeksema. The Wilcox Solar Observatory is currently supported by NASA. 
MESSENGER data were calibrated using measurements from Neutron monitors of the Bartol Research Institute, which are supported by the National Science Foundation. 
Work from D. J. Lawrence is supported by the NASA's MESSENGER Participating Scientist Program through NASA grant NNX08AN30G. All original MESSENGER data reported in this paper are archived by the NASA  Planetary Data System  (\url{http://pdsgeosciences/wustl.edu/missions/messenger/index.htm}). \refboldd{The authors wish to thank Dr. Nina Dresing for her assistance in accessing and using STEREO data.}
The authors gratefully acknowledge the important comments and suggestions provided by the anonymous referee.

\end{acknowledgements}


\bibliographystyle{aa} 
\bibliography{combined}

\begin{thebibliography}{54}
\expandafter\ifx\csname natexlab\endcsname\relax\def\natexlab#1{#1}\fi

\bibitem[{Adriani {et~al.}(2015)Adriani, Barbarino, Bazilevskaya, Bellotti,
  Boezio, Bogomolov, Bongi, Bonvicini, Bottai, Bravar,
  {et~al.}}]{adriani2015pamela}
Adriani, O., Barbarino, G., Bazilevskaya, G., {et~al.} 2015, ApJ, 801, 5

\bibitem[{Afanasiev {et~al.}(2017)Afanasiev, Vainio, Rouillard, Battarbee,
  Aran, \& Zucca}]{Afanasiev2017}
Afanasiev, A., Vainio, R., Rouillard, A.~P., {et~al.} 2017, submitted to A{\&}A

\bibitem[{{Asvestari} {et~al.}(2017){Asvestari}, {Willamo}, {Gil}, {Usoskin},
  {Kovaltsov}, {Mikhailov}, \& {Mayorov}}]{Asvestari2017}
{Asvestari}, E., {Willamo}, T., {Gil}, A., {et~al.} 2017, Advances in Space
  Research, 60, 781

\bibitem[{Battarbee {et~al.}(2018)Battarbee, Dalla, , \& Marsh}]{Battarbee2018}
Battarbee, M., Dalla, S., , \& Marsh, M.~S. 2018, submitted to ApJ

\bibitem[{Battarbee {et~al.}(2017)Battarbee, Dalla, \& Marsh}]{Battarbee2017}
Battarbee, M., Dalla, S., \& Marsh, M.~S. 2017, ApJ, 836, 138

\bibitem[{Berrilli {et~al.}(2014)Berrilli, Casolino, {Del Moro}, {Di Fino},
  Larosa, Narici, Piazzesi, Picozza, Scardigli, Sparvoli, Stangalini, \&
  Zaconte}]{Berrilli2014}
Berrilli, F., Casolino, M., {Del Moro}, D., {et~al.} 2014, Journal of Space
  Weather and Space Climate, 4, A16

\bibitem[{Bieber {et~al.}(2014)Bieber, Evenson, \& Pyle}]{Bieber2014}
Bieber, J.~W., Evenson, P., \& Pyle, R. 2014, {Bartol Research Institute
  neutron monitor data}, http://neutronm.bartol.udel.edu

\bibitem[{Castagnoli \& Lal(1980)}]{castagnoli_lal_1980}
Castagnoli, G. \& Lal, D. 1980, Radiocarbon, 22, 133–158

\bibitem[{Dalla {et~al.}(2013)Dalla, Marsh, Kelly, \& Laitinen}]{Dalla2013}
Dalla, S., Marsh, M.~S., Kelly, J., \& Laitinen, T. 2013, JGR: Space Physics,
  118, 5979

\bibitem[{Dalla {et~al.}(2015)Dalla, Marsh, \& Laitinen}]{Dalla2015}
Dalla, S., Marsh, M.~S., \& Laitinen, T. 2015, ApJ, 808, 62

\bibitem[{Ding {et~al.}(2016)Ding, Jiang, \& Li}]{Ding2016}
Ding, L.-G., Jiang, Y., \& Li, G. 2016, ApJ, 818, 169

\bibitem[{Dresing {et~al.}(2014)Dresing, Cohen, G{\'{o}}mez-Herrero, Heber,
  Klassen, Leske, Mason, Mewaldt, \& von Rosenvinge}]{Dresing2014}
Dresing, N., Cohen, C. M.~S., G{\'{o}}mez-Herrero, R., {et~al.} 2014, Brazilian
  Journal of Physics, 44, 504

\bibitem[{Feldman {et~al.}(2010)Feldman, Lawrence, Goldsten, Gold, Baker,
  Haggerty, Ho, Krucker, Lin, Mewaldt, Murphy, Nittler, Rhodes, Slavin,
  Solomon, Starr, Vilas, \& Vourlidas}]{Feldman2010}
Feldman, W.~C., Lawrence, D.~J., Goldsten, J.~O., {et~al.} 2010, JGR: Space
  Physics, 115, n/a

\bibitem[{Firoz {et~al.}(2015)Firoz, Gan, Li, \&
  Rodr{\'{i}}guez-Pacheco}]{Firoz2015}
Firoz, K.~A., Gan, W.~Q., Li, Y.~P., \& Rodr{\'{i}}guez-Pacheco, J. 2015, Solar
  Physics, 290, 613

\bibitem[{Firoz {et~al.}(2014)Firoz, Zhang, Gan, Li, Rodr{\'{i}}guez-Pacheco,
  Moon, Kudela, Park, \& Dorman}]{Firoz2014}
Firoz, K.~A., Zhang, Q.~M., Gan, W.~Q., {et~al.} 2014, ApJS, 213, 24

\bibitem[{Gopalswamy {et~al.}(2013)Gopalswamy, Xie, Akiyama, Yashiro, Usoskin,
  \& Davila}]{Gopalswamy2013}
Gopalswamy, N., Xie, H., Akiyama, S., {et~al.} 2013, ApJ, 765, L30

\bibitem[{Gopalswamy {et~al.}(2012)Gopalswamy, Xie, Yashiro, Akiyama,
  M{\"{a}}kel{\"{a}}, \& Usoskin}]{Gopalswamy2012}
Gopalswamy, N., Xie, H., Yashiro, S., {et~al.} 2012, Space Sci. Rev., 171, 23

\bibitem[{Hassler {et~al.}(2012)Hassler, Zeitlin, Wimmer-Schweingruber,
  B{\"o}ttcher, Martin, Andrews, B{\"o}hm, Brinza, Bullock, Burmeister,
  {et~al.}}]{hassler2012}
Hassler, D., Zeitlin, C., Wimmer-Schweingruber, R., {et~al.} 2012, Space Sci.
  Rev., 170, 503

\bibitem[{Heber {et~al.}(2013)Heber, Dresing, Dr{\"{o}}ge, G{\'{o}}mez-Herrero,
  Herbst, Kartvykh, Klassen, Kocharov, K{\"{u}}hl, Labrenz, Malandraki, Terasa,
  \& Valtonen}]{Heber2013}
Heber, B., Dresing, N., Dr{\"{o}}ge, W., {et~al.} 2013, Proc. 33rd Internat.
  Cosmic Ray Conf. (Rio de Janeiro), paper 0746

\bibitem[{Hoeksema {et~al.}(1983)Hoeksema, Wilcox, \& Scherrer}]{Hoeksema1983}
Hoeksema, J.~T., Wilcox, J.~M., \& Scherrer, P.~H. 1983, JGR, 88, 9910

\bibitem[{Jian {et~al.}(2013)Jian, Russell, Luhmann, Galvin, \&
  Simunac}]{Jian2013}
Jian, L.~K., Russell, C.~T., Luhmann, J., Galvin, A.~B., \& Simunac, K. D.~C.
  2013, in AIP Conf. Ser. 1539, eds. G. P. Zank, J. Borovsky, R. Bruno, et al.,
  191

\bibitem[{Kocharov {et~al.}(2017)Kocharov, Pohjolainen, Mishev, Reiner, Lee,
  Laitinen, Didkovsky, Pizzo, Kim, Klassen, Karlicky, Cho, Gary, Usoskin,
  Valtonen, \& Vainio}]{Kocharov2017}
Kocharov, L., Pohjolainen, S., Mishev, A., {et~al.} 2017, ApJ, 839, 79

\bibitem[{K{\"{u}}hl {et~al.}(2015)K{\"{u}}hl, Banjac, Dresing,
  Gom{\'{e}}z-Herrero, Heber, Klassen, \& Terasa}]{kuhl2015}
K{\"{u}}hl, P., Banjac, S., Dresing, N., {et~al.} 2015, A{\&}A, 576, A120

\bibitem[{K{\"u}hl {et~al.}(2016)K{\"u}hl, Dresing, Heber, \&
  Klassen}]{Kuhl2016}
K{\"u}hl, P., Dresing, N., Heber, B., \& Klassen, A. 2016, Solar Physics, 292,
  10

\bibitem[{Laitinen {et~al.}(2017)Laitinen, Dalla, \& Marriott}]{Laitinen2017}
Laitinen, T., Dalla, S., \& Marriott, D. 2017, MNRAS, stx1509
  [\eprint[arXiv]{1706.06580}]

\bibitem[{Laitinen {et~al.}(2013)Laitinen, Dalla, \& Marsh}]{Laitinen2013}
Laitinen, T., Dalla, S., \& Marsh, M.~S. 2013, ApJ, 773, L29

\bibitem[{Laitinen {et~al.}(2016)Laitinen, Kopp, Effenberger, Dalla, \&
  Marsh}]{Laitinen2016}
Laitinen, T., Kopp, A., Effenberger, F., Dalla, S., \& Marsh, M.~S. 2016,
  A{\&}A, 591, A18

\bibitem[{Lario {et~al.}(2013)Lario, Aran, G{\'{o}}mez-Herrero, Dresing, Heber,
  Ho, Decker, \& Roelof}]{lario2013}
Lario, D., Aran, A., G{\'{o}}mez-Herrero, R., {et~al.} 2013, ApJ, 767, 41

\bibitem[{Lawrence {et~al.}(2014)Lawrence, Feldman, Goldsten, Peplowski,
  Rodgers, \& Solomon}]{lawrence2014detection}
Lawrence, D.~J., Feldman, W.~C., Goldsten, J.~O., {et~al.} 2014, JGR: Space
  Physics, 119, 5150

\bibitem[{Lawrence {et~al.}(2017)Lawrence, Peplowski, Beck, Feldman, Frank,
  McCoy, Nittler, \& Solomon}]{Lawrence2017}
Lawrence, D.~J., Peplowski, P.~N., Beck, A.~W., {et~al.} 2017, Icarus, 281, 32

\bibitem[{Lawrence {et~al.}(2016)Lawrence, Peplowski, Feldman, Schwadron, \&
  Spence}]{Lawrence2016}
Lawrence, D.~J., Peplowski, P.~N., Feldman, W.~C., Schwadron, N.~A., \& Spence,
  H.~E. 2016, JGR: Space Physics, 121, 7398

\bibitem[{Li {et~al.}(2013)Li, Firoz, Sun, \& Miroshnichenko}]{Li2013}
Li, C., Firoz, K.~A., Sun, L.~P., \& Miroshnichenko, L.~I. 2013, ApJ, 770, 34

\bibitem[{Marsh {et~al.}(2015)Marsh, Dalla, Dierckxsens, Laitinen, \&
  Crosby}]{Marsh2015}
Marsh, M.~S., Dalla, S., Dierckxsens, M., Laitinen, T., \& Crosby, N.~B. 2015,
  Space Weather, 13, 386

\bibitem[{Marsh {et~al.}(2013)Marsh, Dalla, Kelly, \& Laitinen}]{Marsh2013}
Marsh, M.~S., Dalla, S., Kelly, J., \& Laitinen, T. 2013, ApJ, 774, 4

\bibitem[{Masarik \& Reedy(1996)}]{Masarik1996}
Masarik, J. \& Reedy, R.~C. 1996, JGR: Planets, 101, 18891

\bibitem[{McComas {et~al.}(1998)McComas, Bame, Barker, Feldman, Phillips,
  Riley, \& Griffee}]{McComas1998}
McComas, D., Bame, S., Barker, P., {et~al.} 1998, Space Sci. Rev., 86, 563

\bibitem[{McKinney {et~al.}(2006)McKinney, Lawrence, Prettyman, Elphic,
  Feldman, \& Hagerty}]{McKinney2006}
McKinney, G.~W., Lawrence, D.~J., Prettyman, T.~H., {et~al.} 2006, JGR, 111,
  E06004

\bibitem[{Mishev {et~al.}(2014)Mishev, Kocharov, \& Usoskin}]{Mishev2014}
Mishev, A.~L., Kocharov, L.~G., \& Usoskin, I.~G. 2014, JGR: Space Physics,
  119, 670

\bibitem[{M{\"{u}}ller-Mellin {et~al.}(1995)M{\"{u}}ller-Mellin, Kunow,
  Flei{\ss}ner, Pehlke, Rode, R{\"{o}}schmann, Scharmberg, Sierks, Rusznyak,
  Mckenna-Lawlor, Elendt, Sequeiros, Meziat, Sanchez, Medina, del Peral, Witte,
  Marsden, \& Henrion}]{Muller-Mellin1995}
M{\"{u}}ller-Mellin, R., Kunow, H., Flei{\ss}ner, V., {et~al.} 1995, Solar
  Physics, 162, 483

\bibitem[{Nitta {et~al.}(2012)Nitta, Liu, DeRosa, \& Nightingale}]{Nitta2012}
Nitta, N.~V., Liu, Y., DeRosa, M.~L., \& Nightingale, R.~W. 2012, Space Sci.
  Rev., 171, 61

\bibitem[{Papaioannou {et~al.}(2014)Papaioannou, Souvatzoglou, Paschalis,
  Gerontidou, \& Mavromichalaki}]{Papaioannou2014}
Papaioannou, A., Souvatzoglou, G., Paschalis, P., Gerontidou, M., \&
  Mavromichalaki, H. 2014, Solar Physics, 289, 423

\bibitem[{Parker(1958)}]{Parker1958}
Parker, E.~N. 1958, ApJ, 128, 664

\bibitem[{Picozza {et~al.}(2007)Picozza, Galper, Castellini, Adriani, Altamura,
  Ambriola, Barbarino, Basili, Bazilevskaja, Bencardino,
  {et~al.}}]{picozza2007pamela}
Picozza, P., Galper, A., Castellini, G., {et~al.} 2007, Astroparticle physics,
  27, 296

\bibitem[{Posner {et~al.}(2013)Posner, Odstr{\^c}il, MacNeice, Rastaetter,
  Zeitlin, Heber, Elliott, Frahm, Hayes, von Rosenvinge, {et~al.}}]{posner2013}
Posner, A., Odstr{\^c}il, D., MacNeice, P., {et~al.} 2013, Planetary and Space
  Science, 89, 127

\bibitem[{Richardson \& Cane(2010)}]{Richardson2010}
Richardson, I.~G. \& Cane, H.~V. 2010, Solar Physics, 264, 189

\bibitem[{Richardson {et~al.}(2014)Richardson, von Rosenvinge, Cane, Christian,
  Cohen, Labrador, Leske, Mewaldt, Wiedenbeck, \& Stone}]{Richardson2014}
Richardson, I.~G., von Rosenvinge, T.~T., Cane, H.~V., {et~al.} 2014, Solar
  Physics, 289, 3059

\bibitem[{Rouillard {et~al.}(2016)Rouillard, Plotnikov, Pinto, Tirole, Lavarra,
  Zucca, Vainio, Tylka, Vourlidas, Rosa, Linker, Warmuth, Mann, Cohen, \&
  Mewaldt}]{Rouillard2016}
Rouillard, A.~P., Plotnikov, I., Pinto, R.~F., {et~al.} 2016, ApJ, 833, 45

\bibitem[{Ruffolo {et~al.}(2012)Ruffolo, Pianpanit, Matthaeus, \&
  Chuychai}]{Ruffolo2012}
Ruffolo, D., Pianpanit, T., Matthaeus, W.~H., \& Chuychai, P. 2012, ApJ, 747,
  L34

\bibitem[{Shen {et~al.}(2013)Shen, Li, Kong, Hu, Sun, Ding, Chen, Wang, \&
  Xia}]{Shen2013}
Shen, C., Li, G., Kong, X., {et~al.} 2013, ApJ, 763, 114

\bibitem[{Smith {et~al.}(1998)Smith, L'Heureux, Ness, Acu{\~{n}}a, Burlaga, \&
  Scheifele}]{Smith1998}
Smith, C., L'Heureux, J., Ness, N., {et~al.} 1998, Space Sci. Rev., 86, 613

\bibitem[{Torsti {et~al.}(1995)Torsti, Valtonen, Lumme, Peltonen, Eronen,
  Louhola, Riihonen, Schultz, Teittinen, Ahola, Holmlund, Kelh{\"{a}},
  Lepp{\"{a}}l{\"{a}}, Ruuska, \& Str{\"{o}}mmer}]{Torsti1995}
Torsti, J., Valtonen, E., Lumme, M., {et~al.} 1995, Solar Physics, 162, 505

\bibitem[{Turner(2000)}]{Turner2000}
Turner, R. 2000, IEEE Transactions on Plasma Science, 28, 2103

\bibitem[{Winslow {et~al.}(2015)Winslow, Lugaz, Philpott, Schwadron, Farrugia,
  Anderson, \& Smith}]{Winslow2015}
Winslow, R.~M., Lugaz, N., Philpott, L.~C., {et~al.} 2015, JGR: Space Physics,
  120, 6101

\bibitem[{Zeitlin {et~al.}(2013)Zeitlin, Hassler, Cucinotta, Ehresmann,
  Wimmer-Schweingruber, Brinza, Kang, Weigle, B{\"o}ttcher, B{\"o}hm,
  {et~al.}}]{zeitlin2013}
Zeitlin, C., Hassler, D., Cucinotta, F., {et~al.} 2013, Sci, 340, 1080

\end{thebibliography}

\appendix

\section{MESSENGER flux calibration} \label{appendix:mes}

As the MESSENGER NS instrument was not originally designed with SEP proton measurements in mind, calibration and validation of derived fluxes is necessary. 
Absolute flux profiles of protons for the MES $\ge$45 MeV and $\ge$125 MeV energy thresholds were determined using the modelled response and validated with measures of the galactic cosmic ray (GCR) flux. Following \cite{Feldman2010}, the measured count rate, $C$, is related to the proton flux, $F_0$, (in units of protons sec$^{-1}$ sr$^{-1}$ cm$^{-2}$) using $C = GAF_0$, where $G$ is the geometry factor in sr, and $A=100\,\mathrm{cm}^2$ is the detector area. For the two highest energy ranges, the values for $G$ are $G_{\ge 125 \,\mathrm{MeV}} = 1.1 \,\mathrm{sr}$ and $G_{\ge 45 \,\mathrm{MeV}} = 4.25 \,\mathrm{sr}$ \citep{lawrence2014detection}. For borated plastic singles, the geometry factor is approximately $G_\mathrm{singles}\approx 4\pi - 2G_{\ge 45 \,\mathrm{MeV}}$. However, the singles count rate likely contains a substantial fraction of contamination and non-proton background counts, such that its absolute calibration for energetic protons is highly uncertain. The measured count rates \citep{Lawrence2016,Lawrence2017} are converted to fluxes using the above relation with the appropriate geometry factors.  

\begin{figure*}[ht!]
\centering
\includegraphics[trim=0 0 0 0,clip, width=0.45\textwidth]{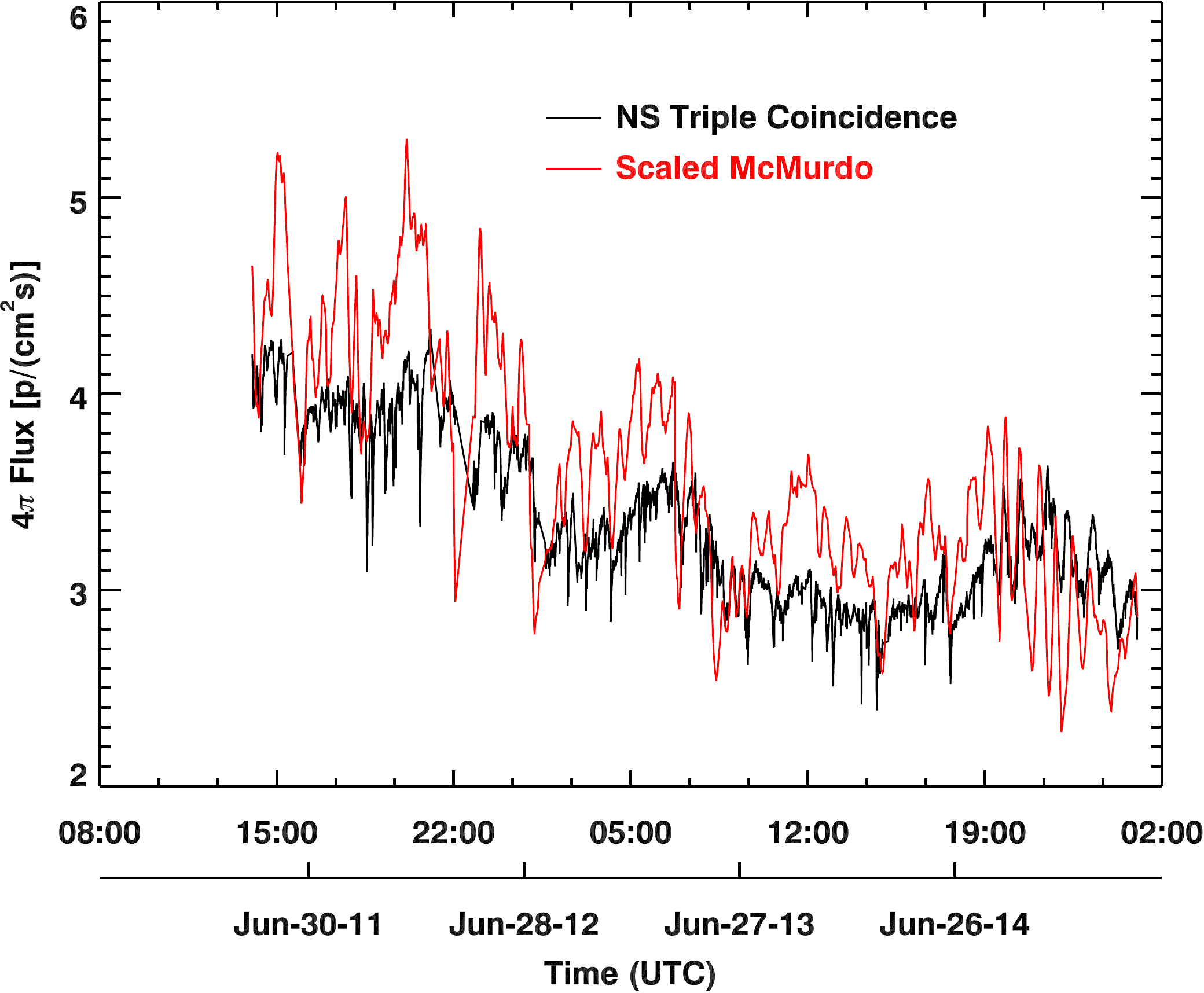}
\includegraphics[trim=0 0 0 0,clip, width=0.45\textwidth]{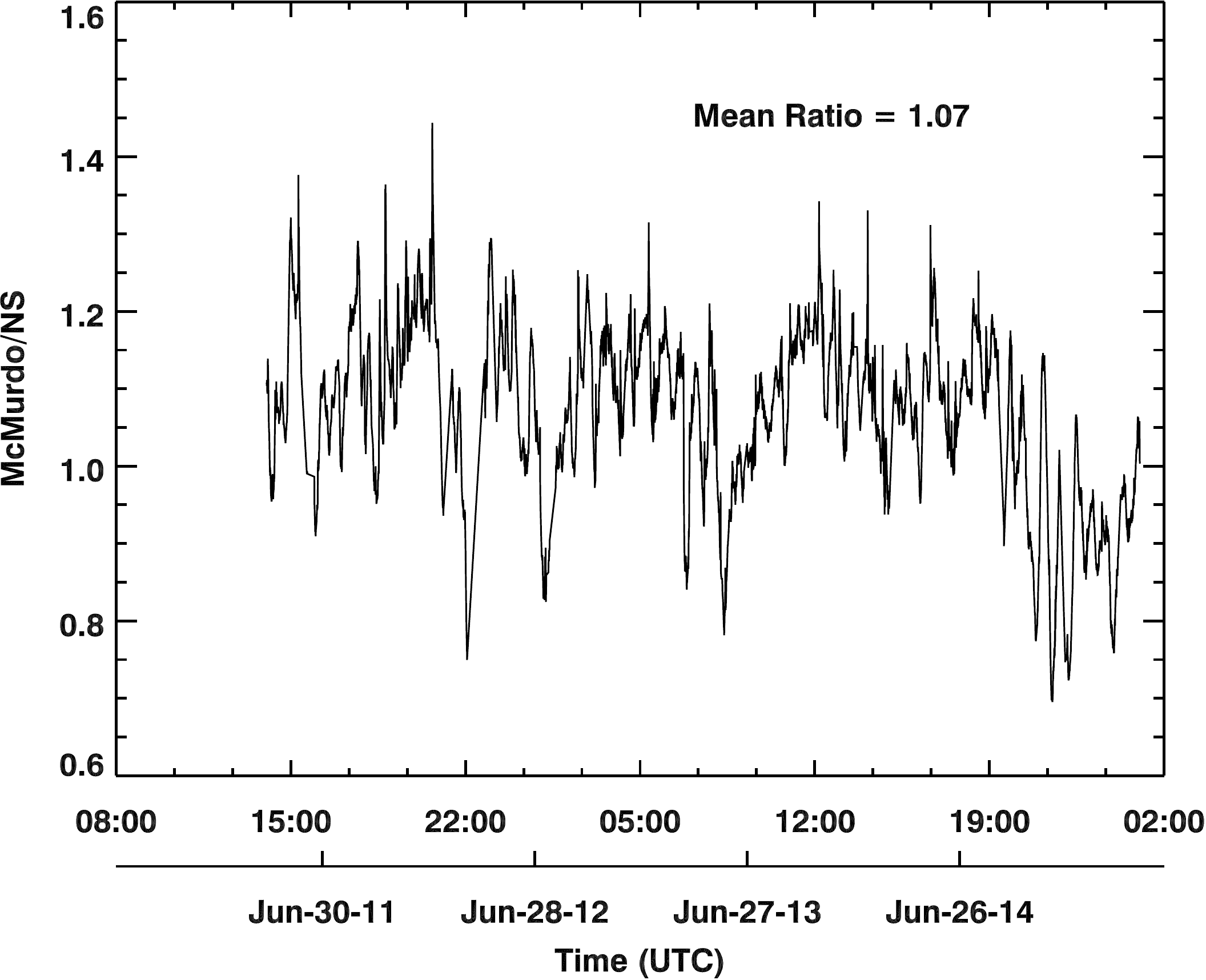}
\includegraphics[trim=0 0 0 0,clip, width=0.45\textwidth]
{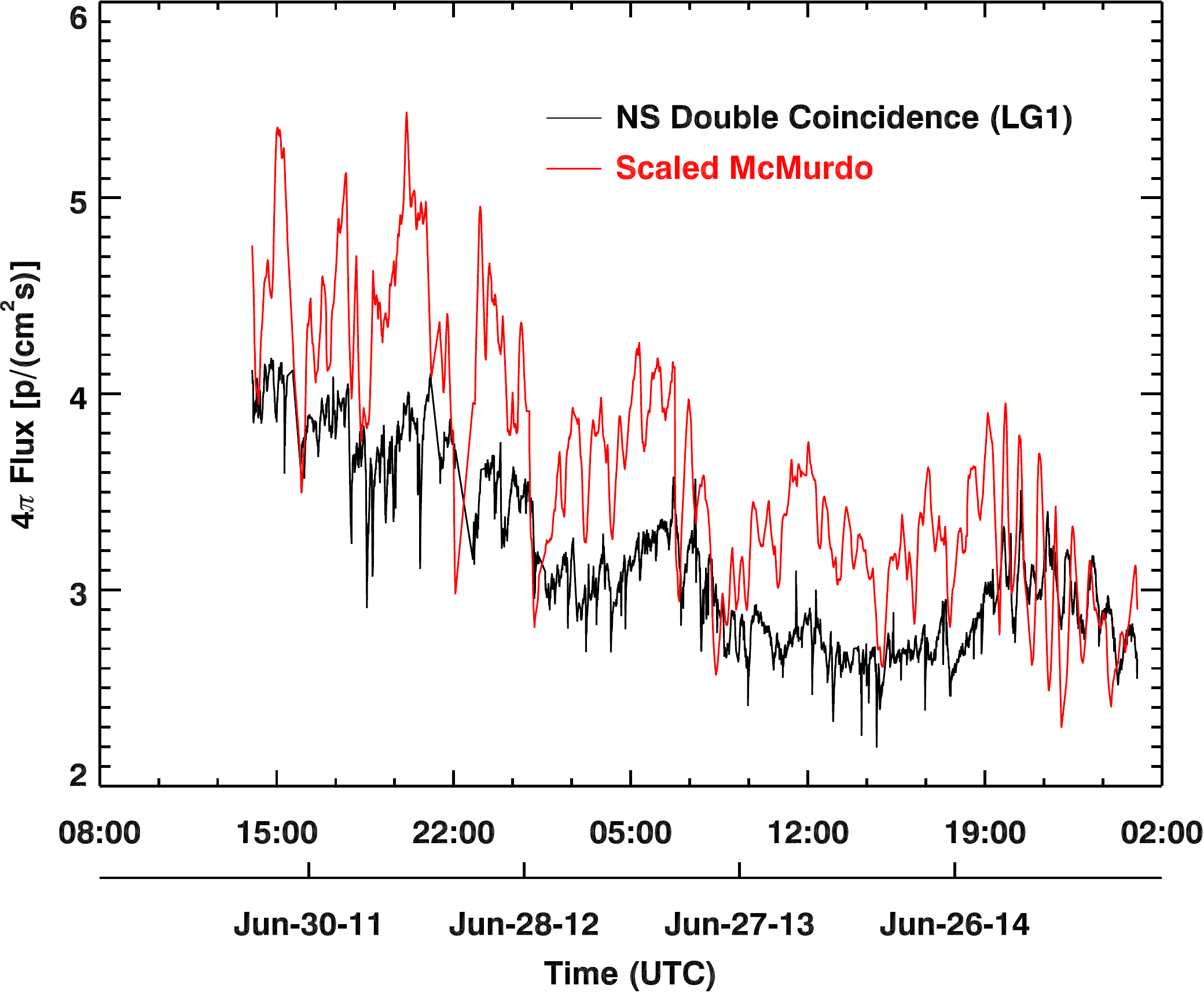}
\includegraphics[trim=0 0 0 0,clip, width=0.45\textwidth]
{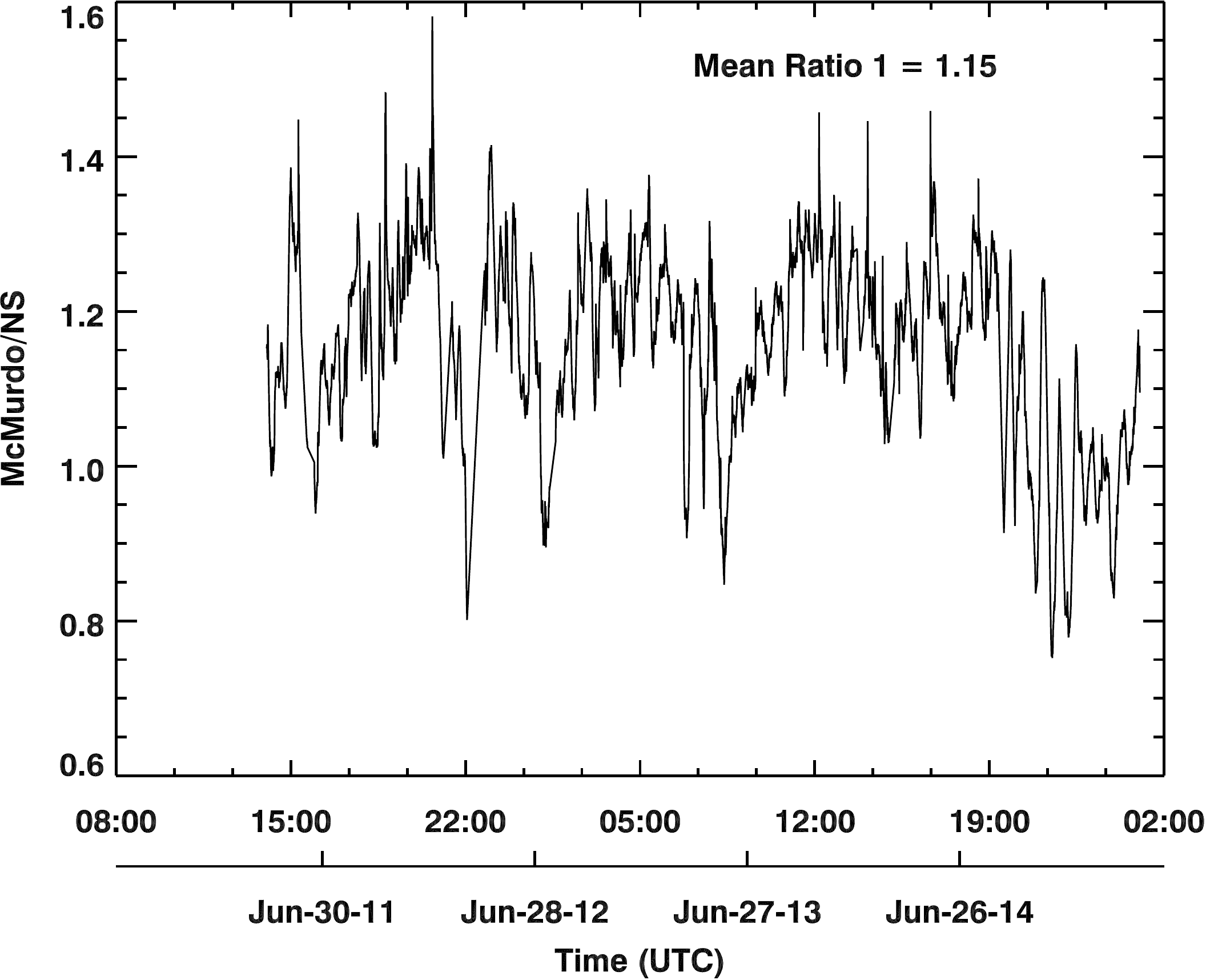}
\includegraphics[trim=0 0 0 0,clip, width=0.45\textwidth]{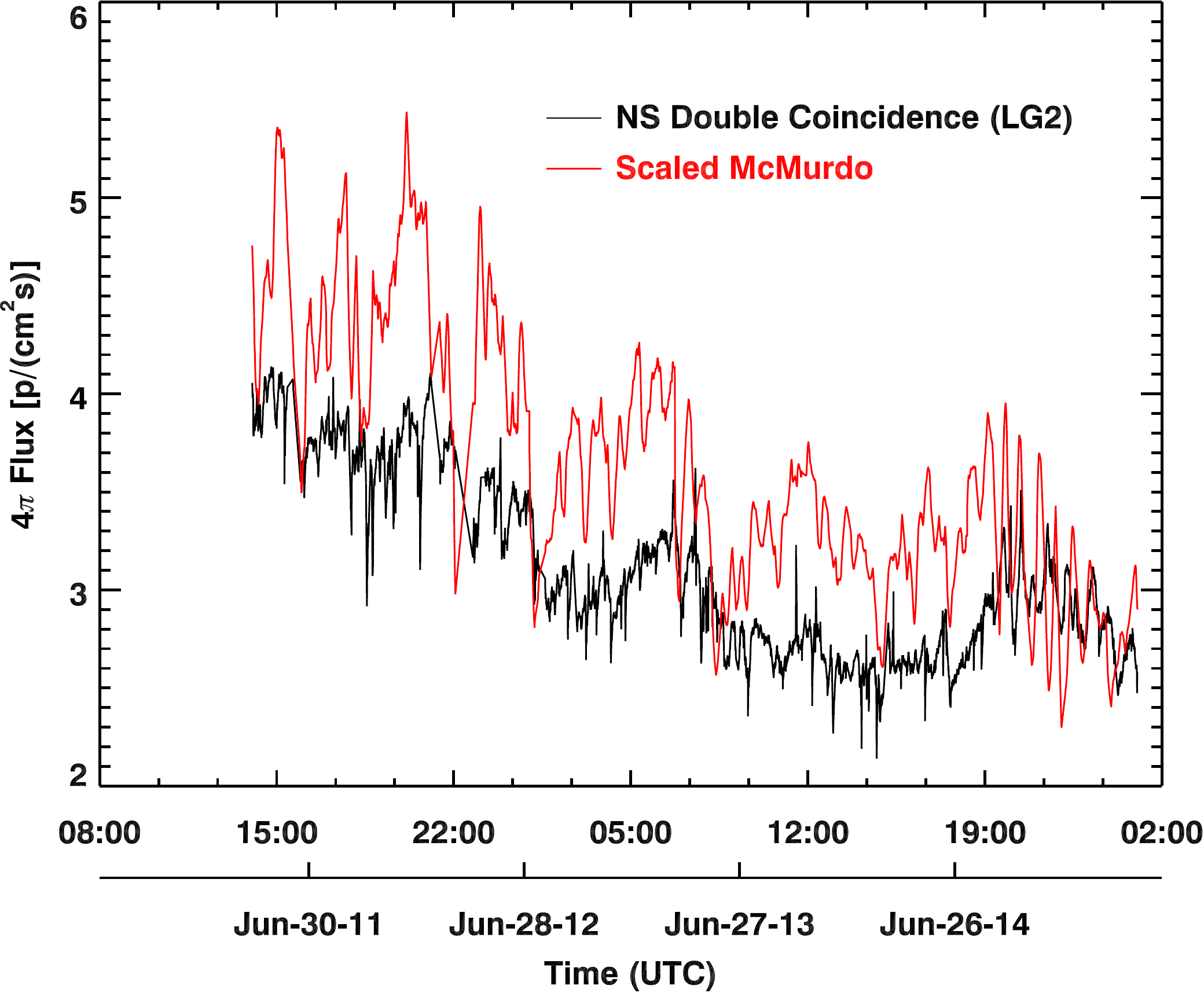}
\includegraphics[trim=0 0 0 0,clip, width=0.45\textwidth]{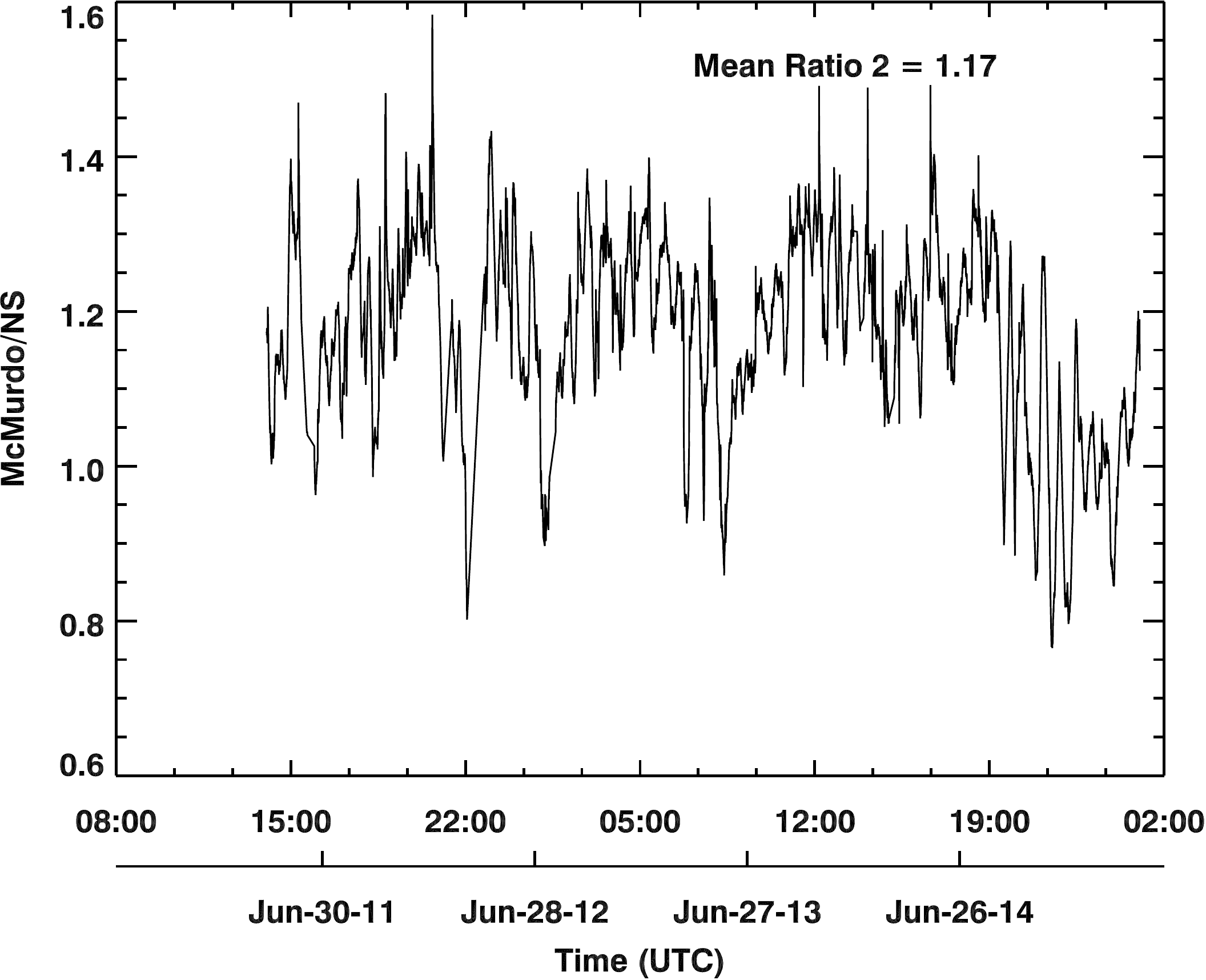}
\caption{Validation for MESSENGER NS observations: GCR flux obtained via MESSENGER NS counts, and from McMurdo neutron monitor observations (left column). Shown are GCR fluxes obtained via NS Triple Coincidence counts (top panel), NS LG1 double coincidence counts (middle panel), and NS LG2 double coincidence counts (bottom panel). The flux ratios (right column) had mean values of 1.07, 1.15, and 1.17, respectively. The time period assessed was March 26th 2011 to April 30th 2015.}\label{fig:NS_NM_validation}
\end{figure*}

The derived fluxes for the $\ge$45 MeV and $\ge$125 MeV thresholds were validated based on a comparison with Earth-based neutron monitor counts that were converted to particle flux using the process given by \cite{McKinney2006}. Specifically, neutron monitor counts from McMurdo \citep{Bieber2014} were empirically converted to a solar modulation parameter, which is used as input to a GCR flux parameterization of \cite{castagnoli_lal_1980} and \cite{Masarik1996}. The total GCR flux accounts for both protons and proton-equivalent alpha particles using the formulation given by \cite{McKinney2006}. When the NS-measured fluxes are compared to the fluxes derived through the neutron monitor data, we find an average absolute agreement of <10\% for the $\ge$125 MeV flux and <20\% for the $\ge$45 MeV flux, which validates the modelled response of \cite{lawrence2014detection}. The flux rates for the time period of March 26th 2011 to April 30th 2015 are plotted in Figure \ref{fig:NS_NM_validation}. 

The mean validation ratios of $1.07$ for triple coincidences, $1.15$ for double coincidence channel LG1 and $1.17$ for double coincidence channel LG2 were applied as correction coefficients to the extracted MES proton fluxes.

\end{document}